\documentclass[preprint]{aastex}
\pdfoutput=0

\newcommand{\vONE}[1]{#1}

\usepackage{amsmath}
\usepackage{natbib}
\usepackage{graphicx}
\usepackage{MyStyle}
\usepackage[
colorlinks=true
]
{hyperref}
\usepackage[all]{hypcap}
\hypersetup{
pdfpagemode = {UseOutlines},
baseurl = {http://www.mpia.de/homes/kuiper},
pdftitle = {Three-dimensional simulation of massive star formation in the disk accretion scenario},
pdfauthor = {R.~Kuiper},
pdfkeywords = {
Accretion disks -
Instabilities -
Stars: formation -
Stars: massive -
Hydrodynamics  -
Methods: numerical}
}

\shortauthors{Kuiper et al.}

\begin{document}

\title{Three-dimensional simulation of massive star formation\\in the disk accretion scenario}

\author{Rolf\ Kuiper\altaffilmark{1,2}}
\email{kuiper@astro.uni-bonn.de}
\author{Hubert\ Klahr\altaffilmark{2}}
\author{Henrik\ Beuther\altaffilmark{2}}
\author{Thomas\ Henning\altaffilmark{2}}
\altaffiltext{1}
{Argelander Institute for Astronomy, Bonn University, Auf dem H\"ugel 71, D-53121 Bonn, Germany} 
\altaffiltext{2}
{Max Planck Institute for Astronomy, K\"onigstuhl 17, D-69117 Heidelberg, Germany}

\begin{abstract}
The most massive stars can form via standard disk accretion - despite of the radiation pressure generated - due to the fact that the massive accretion disk yields a strong anisotropy in the radiation field, releasing most of the radiation pressure perpendicular to the disk accretion flow.
Here, we analyze the self-gravity of the forming circumstellar disk as the potential major driver of the angular momentum transport in such massive disks responsible for the high accretion rates needed for the formation of massive stars.
For this purpose, we perform self-gravity radiation hydrodynamics simulations of the collapse of massive pre-stellar cores.
The formation and evolution of the resulting circumstellar disk is investigated in 
$1.)$ axially symmetric simulations using an $\alpha$-shear-viscosity prescription and
$2.)$ a three-dimensional simulation, in which the angular momentum transport is provided self-consistently by developing gravitational torques in the self-gravitating accretion disk.
The simulation series of different strength of the $\alpha$-viscosity shows that the accretion history of the  forming star is mostly independent of the $\alpha$-viscosity-parameter.
The accretion history of the three-dimensional run driven by self-gravity is more time-dependent than the viscous disk evolution in axial symmetry.
The mean accretion rate, i.e.~the stellar mass growth, is nearly identical to the $\alpha$-viscosity models.
We conclude that the development of gravitational torques in self-gravitating disks around forming massive stars provides a self-consistent mechanism to efficiently transport the angular momentum to outer disk radii.
Also the formation of the most massive stars can therefore be understood in the standard accretion disk scenario.
\end{abstract}


\maketitle

\section{Introduction}
\label{sect:Introduction}
The formation of circumstellar disks is a natural outcome of pre-stellar core collapse simulations regarding the formation of stars due to angular momentum conservation.
In \citet{Kuiper:2010p17191} we demonstrated the possibility, how to overcome the well-known radiation pressure barrier in the formation of massive stars via disk accretion.
These simulations were performed in axial symmetry, in which the angular momentum transport in the circumstellar disk relies on a standard $\alpha$-shear-viscosity prescription.
In the three-dimensional simulation presented herein, we are now able to check these presumptions by studying the accretion rate driven self-consistently by gravitational torques, which develop from the non-axially symmetric disk structure.

Once a nearly Keplerian disk has formed, further accretion onto the proto-star is only possible if angular momentum is either removed from the disk or efficiently transported to outer disk radii.
Proto-stellar jets, outflows and disk winds allow to remove a substantial fraction of the angular momentum from the star-disk system.
Developing 
convective \citep{Kley:1993p15156, Lin:1993p15154}, 
radiation hydrodynamical \citep{Klahr:2003p2794}, 
magneto-rotational \citep{Balbus:1991p3799, Hawley:1991p11900} and 
self-gravitating instabilities \citep{Cassen:1981p15165, Anthony:1988p14843, Lin:1990p14979, Yang:1991p11485, Papaloizou:1991p14835, Tomley:1991p14849, Heemskerk:1992p15176, Tomley:1994p14850, Laughlin:1994p11930, Miyama:1994p15169, Papaloizou:1995p15024, Laughlin:1996p15224, Laughlin:1996p15239, Bate:1998p15242, Yorke:1999p1156, Vorobyov:2006p16253, Vorobyov:2007p15725, Vorobyov:2010p16256} 
in the accretion disk will transfer angular momentum to outer radii.
Turbulent viscosity as the main mechanism for angular momentum transport in disk has been studied in axially symmetric simulations, including radiation transport \citep{Ruden:1991p15061, Kley:1992p15153,  Kley:1993p15155, Rozyczka:1994p15157}, in non-axially symmetric two-dimensional simulations in the thin disk approximation \citep{Vorobyov:2009p16255} as well as in three-dimensional adiabatic disk simulations \citep{Kley:1993p15156, Lin:1993p15154}.

Self-gravity has been shown to be a major driver of angular momentum transport via developing gravitational torques.
The numerous studies on self-gravity mentioned above involve a variety of different approaches and methods:
When starting from a collapse of a (1 to 10~$\mbox{M}_\odot$) pre-stellar core to compute the formation of the accretion disk consistently in the model,
semi-analytical work \citep{Lin:1990p14979},
two-dimensional grid based hydrodynamics simulations \citep{Laughlin:1994p11930, Yorke:1999p1156}, including radiation transport, and follow-up three-dimensional SPH disk simulations \citet{Laughlin:1994p11930}, 
as well as
three-dimensional SPH simulations of the collapse \citep{Bate:1998p15242}
have been considered.

Several methods have been used when studying the angular momentum transport with a disk model as an initial condition:
The first simulations performed \citep{Cassen:1981p15165, Anthony:1988p14843, Tomley:1991p14849, Tomley:1994p14850} were two-dimensional N-body simulations.
\citet{Adams:1989p14769} and \citet{Laughlin:1996p15224} did semi-analytical work.
\citet{Papaloizou:1991p14835},
\citet{Heemskerk:1992p15176}, and
\citet{Laughlin:1996p15239} used
two-dimensional grid based hydrodynamics with a polytropic equation of state.
The first three-dimensional grid based hydrodynamics were performed by \citet{Yang:1991p11485},
\citet{Miyama:1994p15169} performed two-dimensional SPH simulations.
Self-gravitating disks were studied with and without the effect of an additionally small amount of viscosity, using a barotropic equation of state and taking into account the vertical disk structure \citep{Papaloizou:1995p15024}.

Several reviews outlined the importance of the ongoing research on the angular momentum transport in disks.
The basic mechanism for angular momentum transport in disks due to gravitational torques has already been reviewed by \citet{Toomre:1977p14986}.
\citet{Tscharnuter:1993p2109} overview theoretical core collapse models with a focus on the angular momentum transport by either turbulent viscosity (two-dimensional models) as well as gravitational torques (three-dimensional models).
\citet{Bodenheimer:1995p11927} present a review on observations and angular momentum transport mechanisms.
\citet{Papaloizou:1995p15025, Lin:1996p15022} comprise angular momentum transfer processes in a variety of astrophysical contexts. 
A more recent review on angular momentum transport in accretion disks is presented by \citet{Balbus:2003p11906}
and an overview of standard accretion disk theory in general is given in \citet{Lodato:2008p10897}.

The studies mentioned above focus on the collapse of a pre-stellar core similar to the solar nebula yielding the formation of a sun-like star. 
Here, we present an investigation of the angular momentum transport in accretion disks around massive stars.
We follow the collapse of massive pre-stellar cores to compute consistently the formation of circumstellar accretion disks around massive stars.
In the case of axially symmetric disks, we follow their evolution through the whole stellar accretion epoch.
For this purpose, we make use of our recently developed self-gravity radiation hydrodynamics code including frequency dependent stellar luminosity feedback.
The problem of angular momentum transport in such accretion disks around massive stars is studied both in axially symmetric runs by using an $\alpha$-viscosity parameterization and in three-dimensional runs, in which gravitational torques develop from the non-axisymmetric structure and drive the accretion flow through the self-gravitating disk.
 
Observations of this early phase of massive star formation suffer mainly from the high extinction of the stellar environment as well as the long distance to massive star forming regions.
Recent and upcoming future generations of space telescopes and interferometric systems such as the Herschel Space Observatory, the James Webb Space Telescope (JWST) and the Atacama Large Millimeter Array (ALMA) will provide a deeper insight into the mechanisms of massive accretion disks.
A list of disk candidates around massive proto-stars is already published in \citet{Cesaroni:2007p1870}.
Recently, \citet{Beuther:2009p12913} conclude from a study of twelve disk candidates that the inner accretion disks, whose existence are assumed due to the high collimation degree of observed jets and outflows, have radii below 1000 AU and are fed by the in-falling outer envelope.
So far, these observations fully match into the disk accretion scenario for the formation of massive stars, which we propose in this study.

\section{Method}
\label{sect:Method}
\subsection{Equations}
\label{sect:Equations}
To follow the motion of the gas, we solve the equations of compressible self-gravity radiation hydrodynamics:
\begin{eqnarray}
\label{eq:Hydrodynamics_Density}
\partial_t \rho + \vec{\nabla} \cdot \left(\rho \vec{u}\right) &=& 0 \\
\label{eq:Hydrodynamics_Momentum}
\partial_t \left(\rho \vec{u}\right) + \vec{\nabla} \cdot \left(\rho \vec{u} \vec{u} + P\right) &=& \vec{f}
\\
\label{eq:Hydrodynamics_Energy}
\partial_t E + \vec{\nabla} \cdot \left(\left(E + P\right) \vec{u}\right) 
&=& 
\vec{u} \cdot \vec{f}
- \vec{\nabla} \cdot \vec{F}_\mathrm{tot} \\
\label{eq:Poisson}
\vec{\nabla}^2 \Phi &=& 4 \pi G \rho -2 G M_* / r^3 \\
\label{eq:Radiation_Diffusion}
\partial_t E_\mathrm{R} + f_\mathrm{c} \vec{\nabla} \cdot \vec{F} &=& - f_\mathrm{c} \left(\vec{\nabla} \cdot \vec{F}_* - Q^+ \right) \\
\label{eq:Radiation_Irradiation}
\vec{F}_*\left(\nu, r\right)/\vec{F}_*\left(\nu, R_*\right) &=& 
\left(R_*/r\right)^2
\exp\left(-\tau\left(\nu, r\right)\right)
\end{eqnarray}
Eqs.~\eqref{eq:Hydrodynamics_Density} to \eqref{eq:Hydrodynamics_Energy} are the conservation equations of hydrodynamics.
The evolution of the gas density $\rho$, velocity $\vec{u}$, pressure $P$, and total energy density $E$ is computed using the open source magneto-hydrodynamics code Pluto3 \citep{Mignone:2007p3421}, including full tensor viscosity.
The  force density vector
\begin{equation}
\label{eq:Forces}
\vec{f} = - \rho \vec{\nabla} \Phi 
+ \vec{\nabla} \Pi 
- \lambda \vec{\nabla} E_\mathrm{R}
- \vec{\nabla} \cdot (\vec{F}_*/c) \vec{e}_r
\end{equation}
includes the additionally considered physics to the equations of gas dynamics such as gravity, shear viscosity (in 2D only), and radiation transport. 
To close the system of the hydrodynamical Eqs.~\eqref{eq:Hydrodynamics_Density} to \eqref{eq:Hydrodynamics_Energy}, we use an ideal gas equation of state
\begin{equation}
P = \left(\gamma - 1\right) E_\mathrm{int},
\end{equation}
which relates the gas pressure $P$ to the internal energy 
$E_\mathrm{int} = E - 0.5 \rho u^2$.
The adiabatic index $\gamma$ is set to $5/3$.

Eq.~\eqref{eq:Poisson} is Poisson's equation with the gravitational potential $\Phi$ and the gravity constant $G$.
$M_*$ denotes the central stellar mass.
Our implementation of Poisson's Eq.~\eqref{eq:Poisson} via a diffusion Ansatz is presented in \citet{Kuiper:2010p17191}.

Eqs.~\eqref{eq:Radiation_Diffusion} and \eqref{eq:Radiation_Irradiation} describe the radiation transport. 
In our recently developed frequency dependent radiation transport method \citep{Kuiper:2010p12874} the flux of the total radiation energy density $\vec{F}_\mathrm{tot}$ is split into the flux of the frequency averaged (hereafter gray) thermal radiation energy density $\vec{F}$ and the flux $\vec{F}_*(\nu,r)$ of the frequency dependent stellar irradiation.
Eq.~\eqref{eq:Radiation_Diffusion} denotes the evolution of the thermal radiation energy density $E_\mathrm{R}$.
The factor $f_\mathrm{c} = \left(c_\mathrm{V} \rho / 4 a T^3 + 1 \right)^{-1}$ depends only on the ratio of internal to radiation energy and contains the specific heat capacity $c_\mathrm{V}$ and the radiation constant $a$.
The source term $Q^+$ depends on the physics included and contains any additional energy source such as hydrodynamical compression $-P \vec{\nabla} \cdot \vec{u}$ and viscous heating.
We solve Eq.~\eqref{eq:Radiation_Diffusion} by using the so-called flux-limited diffusion approximation (hereafter FLD), in which the flux is set proportional to the gradient of the radiation energy density ($\vec{F} = - D \vec{\nabla} E_\mathrm{R}$).
The diffusion constant $D = \lambda c / \rho \kappa_\mathrm{R}$ depends on the flux limiter $\lambda$, the speed of light in vacuum $c$, and the Rosseland mean opacity $\kappa_\mathrm{R}$.
We use the flux limiter by \citep{Levermore:1981p57} and neglect scattering.

Eq.~\eqref{eq:Radiation_Irradiation} calculates the flux of the frequency dependent stellar irradiation in a ray-tracing step.
The first factor on the right hand side describes the geometrical decrease of the flux proportional to $r^{-2}$.
The second factor describes the absorption of the stellar light as a function of the optical depth $\tau(\nu,r)=\kappa(\nu) \rho(r) r$ depending on the frequency dependent mass absorption coefficients $\kappa(\nu)$.
For this purpose, we use tabulated dust opacities by \citet{Laor:1993p736}, including 79 frequency bins, and calculate the local evaporation temperature of the dust grains by using the formula of \citet{Isella:2005p3014}.
The flux at the inner radial boundary is given by the luminosity $L_*$, temperature $T_*$, and radius $R_*$ of the forming star.
For this purpose,  we use tabulated stellar evolutionary tracks for accreting high-mass stars, recently calculated by \citet{Hosokawa:2009p12591}.
The gas and dust temperature $T$ is finally calculated in equilibrium with the combined stellar irradiation and thermal radiation field
\begin{equation}
a T^4 = E_\mathrm{R} + \frac{\kappa\left(\nu\right)}{\kappa_\mathrm{P}(T)} \frac{|\vec{F}_*|}{c}
\end{equation}
with the Planck mean opacities $\kappa_\mathrm{P}$.

Numerical details, test cases, including a comparison of gray and frequency dependent irradiation, as well as performance studies of our recently developed hybrid radiation transport scheme are summarized by \citet{Kuiper:2010p12874}.
The viscosity prescription as well as the tabulated dust and stellar evolution model are presented in \citet{Kuiper:2010p17191}.

To limit the range of densities, the so-called floor value of the density is chosen to be $\rho_0 = 10^{-21} \rhocgs$.
This floor value occurs during the simulations only in regions where the radiation pressure driven outflow is depleting the density of the corresponding grid cells in the radially outward direction. 
Thus, the choice of the floor value does not influence the level of accretion onto the newly forming star we investigate.

To control the spatial resolution, which is necessary to resolve the physics of self-gravity correctly, e.g.~preventing artificial fragmentation, we monitor the so-called Truelove criterion, derived in \citet{Truelove:1997p10742}.

In Eq.~\eqref{eq:Forces} the viscosity term $\vec{\nabla} \Pi$ is added to the conservation equations of hydrodynamics Eqs.~\eqref{eq:Hydrodynamics_Momentum} and
\eqref{eq:Hydrodynamics_Energy} in case of axially symmetric, two-dimensional simulations only.
The components of the viscous stress tensor $\Pi$ are given (in Cartesian coordinates) by
\begin{equation}
\Pi_{ij} = \eta \left(\partial_j u_i + \partial_i u_j - \frac{2}{3} \delta_{ij} \partial_k u_k \right) 
+ \eta_\mathrm{b} \delta_{ij} \partial_k u_k
\end{equation}
with 
the shear viscosity $\eta$,
the bulk viscosity $\eta_\mathrm{b}$,
and the Kronecker symbol $\delta_{ij}$.
We assume for the bulk viscosity $\eta_\mathrm{b} = 0$.
The (shear) viscosity is described via the so-called $\alpha$-parameterization of
\citet{Shakura:1973p3060} and is computed via
\begin{equation}
\eta = \rho ~ \alpha ~ \Omega_\mathrm{K}(r) ~ R^2 \left(H/R\right)^2
\end{equation}
with the Keplerian angular velocity
\begin{equation}
\Omega_\mathrm{K}(r) = \sqrt{\frac{G M(r)}{r^3}},
\end{equation}
the cylindrical radius $R = r ~ \sin(\theta)$,
and the  pressure scale height $H$.
$M(r)$ denotes the mass inside the radius $r$ and is given by the sum of the central stellar mass and the integral over the gas density in the computational domain
\begin{equation}
\label{eq:includedmass}
M(r) = M_* + 2\pi ~ \int_0^r dr \int_0^\pi d\theta ~ \rho(r,\theta) ~ r^2 \sin(\theta).
\end{equation}
\vONE{
In fact, the resulting rotation profiles of the forming disks in our simulations are very close to Keplerian ($(v - v_\mathrm{K})/v_\mathrm{K} < 2\%$).
Of course, the actual outer edge $r_\mathrm{disk}$ of the accretion disk, in which Keplerian equilibrium is nearly satisfied, expands in time.
}

If the viscosity is an effect of turbulent transport of angular momentum, it is observed that the strength of the stresses is proportional to the thermal pressure. 
This is the fundamental assumption of the $\alpha$-Ansatz by \citet{Shakura:1973p3060}.
This relation holds because hotter and thicker disks can support higher levels of turbulence.
The situation is reversed for self-gravitating disks.
Here, hot disks are usually Toomre stable and will not produce gravito-turbulence. 
On the contrary, the disks will cool down to the marginally unstable Toomre values and establish a turbulent state where the level of turbulence is set by the equilibrium of energy release and radiative cooling \citep{Gammie:2001p14271}.
For that reason, we choose a viscosity prescription independent on the actual disk temperature (e.g.~a fixed $H/R$ ratio of $0.1$) but only on the local mean (Keplerian) rotation profile.
This way, we ensure that cool and thin disks can obtain the high viscosity values they deserve.
Our Ansatz is equivalent to the so-called $\beta$-viscosity Ansatz for self-gravitating disks by \citet{Duschl:2000p14177}, which is also independent on temperature.
A detailed derivation of the viscosity prescription is given in \citet{Kuiper:2010p17191}.

\subsection{Numerical configuration}
The simulations are performed on a time independent grid in spherical coordinates with a logarithmically stretched radial coordinate axis.
The outer core radius is fixed to $r_\mathrm{max} = 0.1$~pc. 
The inner core radius is fixed to $r_\mathrm{min} = 10$~AU. 
The accurate size of this inner sink cell was determined in a parameter scan presented in \citet{Kuiper:2010p17191}, Sect.~5.1.
The polar angle extents from $0^\circ$ to $90^\circ$ assuming midplane symmetry.
In the three-dimensional run, the azimuthal angle covers the full domain from $0^\circ$ to $360^\circ$.
The grid consists of 64 x 16 x 64 grid cells, i.e.~the highest resolution of the non-uniform grid is chosen to be 
\begin{equation}
\Delta r \mbox{ x } r\Delta{\theta} \mbox{ x } r \Delta{\phi} \sin(\theta) = 1.27 \mbox { x } 1.04 \mbox { x } 1.04 \mbox{ AU}^3
\end{equation}
in the midplane ($\theta = 90^\circ$) around the forming massive star.
The resolution decreases logarithmically in the radial outward direction proportional to the radius.
The radially inner and outer boundary of the computational domain are semi-permeable walls, i.e.~the gas is allowed to leave the computational domain (due to radiative forces) but cannot enter it.
This outer boundary condition allows that the mass reservoir for stellar accretion can be controlled by the initial choice of the mass of the pre-stellar core.

The Pluto code uses high-order Godunov solver methods for computing the hydrodynamics, 
i.e.~it uses a shock capturing Riemann solver within a conservative finite volume scheme. 
The numerical configuration of our simulations makes use of a Strang operator splitting scheme for the different dimensions \citep{Strang:1968p6618}.
Our default configuration consists further of a Harten-Lax-Van~Leer Riemann solver and a so-called ``minmod'' flux limiter using piecewise linear interpolation and a Runge-Kutta~2 time integration, also known as the predictor-corrector-method; for comparison please see \citet{vanLeer:1979p5193}. 
Therefore the total difference scheme is accurate to second order in time and space.

The internal iterations of the implicit solver for the FLD Eq.~\eqref{eq:Radiation_Diffusion} is stopped at an accuracy of the resulting temperature distribution of $\Delta T / T \le 10^{-3}$ or $\Delta T \le 0.1$~K.
The internal iterations of the implicit solver for Poisson's Eq.~\eqref{eq:Poisson} is stopped at an accuracy of the resulting gravitational potential of $\Delta \Phi / \Phi \le 10^{-5}$.

\subsection{Initial conditions}
\label{sect:InitialConditions}
The basic physical initial conditions are very similar to the one used by \citet{Yorke:2002p1}.
We start from a cold ($T_0 = 20 \mbox{ K}$) pre-stellar core of gas and dust.
The initial dust to gas mass ratio is chosen to be $M_\mathrm{dust} / M_\mathrm{gas} = 1\%$.
The model describes a so-called quiescent collapse scenario without initial turbulent motion ($\vec{u}_r = \vec{u}_\theta = 0$).
The core is initially in slow solid-body rotation $\left(|\vec{u}_\phi| / R = \Omega_0 = 5*10^{-13} \mbox{ Hz}\right)$.
In the axially symmetric runs, the total mass $M_\mathrm{core}$ in the computational domain is 60 or 120\Msol (17.2 or 48.6 $M_\mathrm{J}$, respectively) and the value for the physical $\alpha$-viscosity varies from $\alpha = 0$ up to $\alpha = 1$.
In the three-dimensional run, the total mass $M_\mathrm{core}$ in the computational domain is chosen to be 120\Msol  (48.6~$M_\mathrm{J}$).
An overview of the runs performed is given in Table~\ref{tab:Runs}.
For a more comprehensive parameter scan of the initial conditions of the axially symmetric core collapse simulations, please see \citet{Kuiper:2010p17191}.
\begin{deluxetable}{l c c c c c}
\tablecaption{Overview of simulations presented.
\label{tab:Runs}}
\tablehead{
\colhead{Label} &
\colhead{Dimension}  &
\colhead{$M_\mathrm{core}~[\mbox{M}_\odot]$} & 
\colhead{$\alpha$} &
\colhead{$t_\mathrm{end}$ [kyr]} &
\colhead{$M_*(t_\mathrm{end})~[\mbox{M}_\odot]$}
}
\startdata 
2D-60Msol-alpha1.0	& 2D &   60 & 1.0   & $  449.0$		& 28.7\\
2D-60Msol-alpha0.3	& 2D &   60 & 0.3   & $  939^{**}$	& 28.2\\
2D-60Msol-alpha0.1	& 2D &   60 & 0.1   & $  240.7$		& 21.7\\
2D-60Msol-alpha0.05	& 2D &   60 & 0.05 & 126.7		&   16.8\\
2D-60Msol-alpha0.03	& 2D &   60 & 0.03 & 	48.0	&  12.6 \\
2D-60Msol-alpha0.02	& 2D &   60 & 0.02 & 	48.3	&  12.1 \\
2D-60Msol-alpha0.01	& 2D &   60 & 0.01 & $    53.4$		&   11.4\\
2D-60Msol-alpha0	& 2D &   60 & 0      & $    24.0$		&   8.8\\
\hline
2D-120Msol-alpha1.0		& 2D & 120 & 1.0	& 14.5	& 23.2 \\
2D-120Msol-alpha0.3		& 2D & 120 & 0.3	& $  489.0^{**}$	& 56.5\\
2D-120Msol-alpha0.1		& 2D & 120 & 0.1	& 38.7	& 25.6\\
2D-120Msol-alpha0.03	& 2D & 120 & 0.03	& 34.6	& 24.7\\
2D-120Msol-alpha0.01	& 2D & 120 & 0.01	& 11.3	& 18.7\\
2D-120Msol-alpha0.003	& 2D & 120 & 0.003	& 10.8	& 18.2\\
2D-120Msol-alpha0.001	& 2D & 120 & 0.001	& 9.3	& 18.1\\
2D-120Msol-alpha0		& 2D & 120 & 0		& $    8.2$		& 18.0\\
3D-120Msol			& 3D & 120 & 0		& $    11.8$		& 24.5\\
\enddata 
\tablecomments{
The runs differ in their dimension, their initial pre-stellar core mass $M_\mathrm{core}$, and their value of the $\alpha$-viscosity-parameter.
The period $t_\mathrm{end}$ of evolution simulated and the corresponding stellar mass $M_*$ at $t_\mathrm{end}$ are given.
A ’**’ denotes that the computation has been stopped at the point in time when no mass is left in the computational domain.
}
\end{deluxetable} 

\clearpage
\section{Results}
\label{sect:Results}
\subsection{Viscous disks (2D)}
\label{sect:2D}
Developing convective \citep{Kley:1993p15156, Lin:1993p15154}, radiation hydrodynamical \citep{Klahr:2003p2794}, magneto-rotational \citep{Balbus:1991p3799, Hawley:1991p11900, Flock:2010p15350, Dzyurkevich:2010p15349} and self-gravitating instabilities \citep{Yang:1991p11485, Laughlin:1994p11930, Bodenheimer:1995p11927, Vorobyov:2006p16253, Vorobyov:2007p15725, Vorobyov:2010p16256} in the accretion disk will transfer angular momentum to outer radii.
To mimic the effect of this angular momentum transport, we made use of the $\alpha$-viscosity model by \citet{Shakura:1973p3060}.
The differential rotation in the circumstellar disk yields a shear viscosity, transporting angular momentum to outer disk radii, while enhancing the stellar accretion rate.

\begin{figure*}[p]
\begin{center}
\hspace{8mm}\includegraphics[width=1.03\FigureWidth]{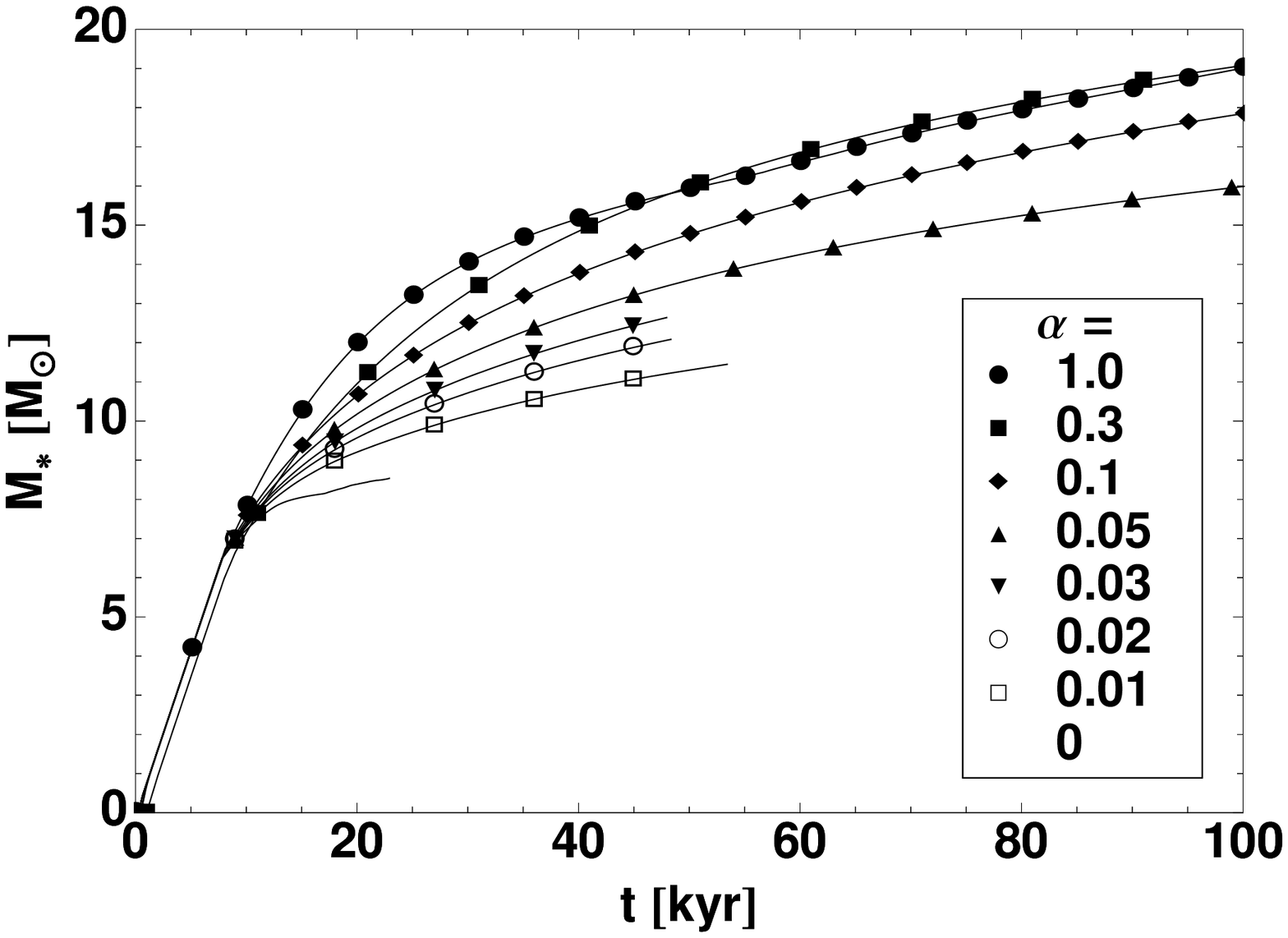}

\includegraphics[width=\FigureWidth]{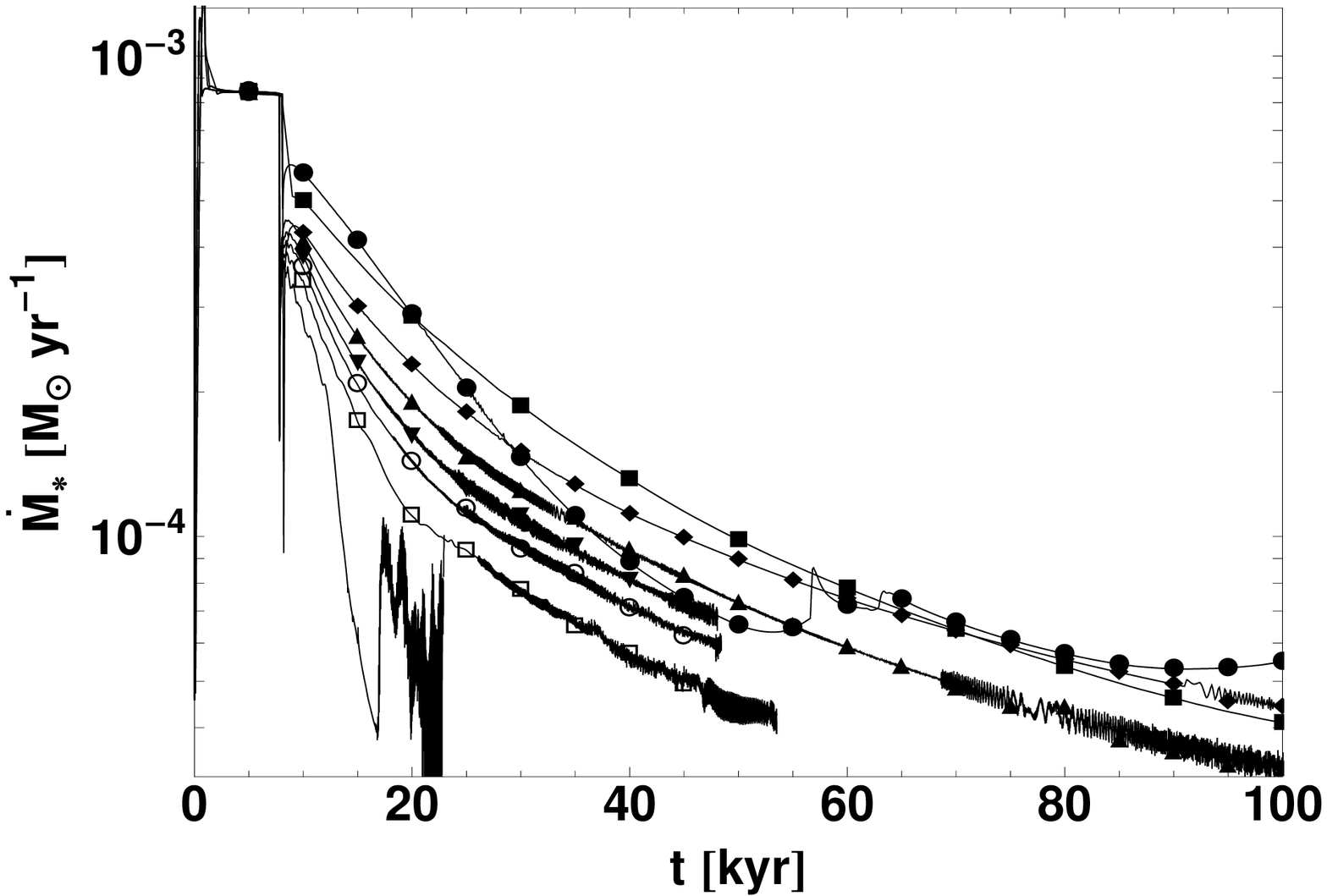}
\end{center}
\caption{
Stellar mass $M_*$ (upper panel) and accretion rate $\dot{M}_*$ (lower panel) as a function of time $t$ for different values of the $\alpha$-shear parameter during the collapse of an axially symmetric 60\Msol pre-stellar core.
\newline
Remark:
The $\alpha$-values mentioned here are associated with a constant aspect ratio $H/R = 0.1$ of the circumstellar disk as described in Sect.~\ref{sect:Equations}. 
Essentially, the $H/R$ ratio computed from the simulation data depends on the radius and increases slightly from 0.05 at the inner disk rim up to 0.15 at the outer disk radius at $r \sim 200$~AU.
}
\label{fig:2D}
\end{figure*}

To study the influence of the value of the $\alpha$-shear-parameter on the resulting stellar accretion rate, we first perform several simulations of an axially symmetric 60\Msol pre-stellar core collapse with varying normalization values for the physical $\alpha$-viscosity.
The resulting stellar mass and the accretion rate as a function of time is shown in Fig.~\ref{fig:2D}.

\clearpage
\subsection{Self-gravitating disks (3D)}
\label{sect:3D}
\begin{figure*}[p]
\begin{center}
\hspace{9mm}\includegraphics[width=0.9\FigureWidth]{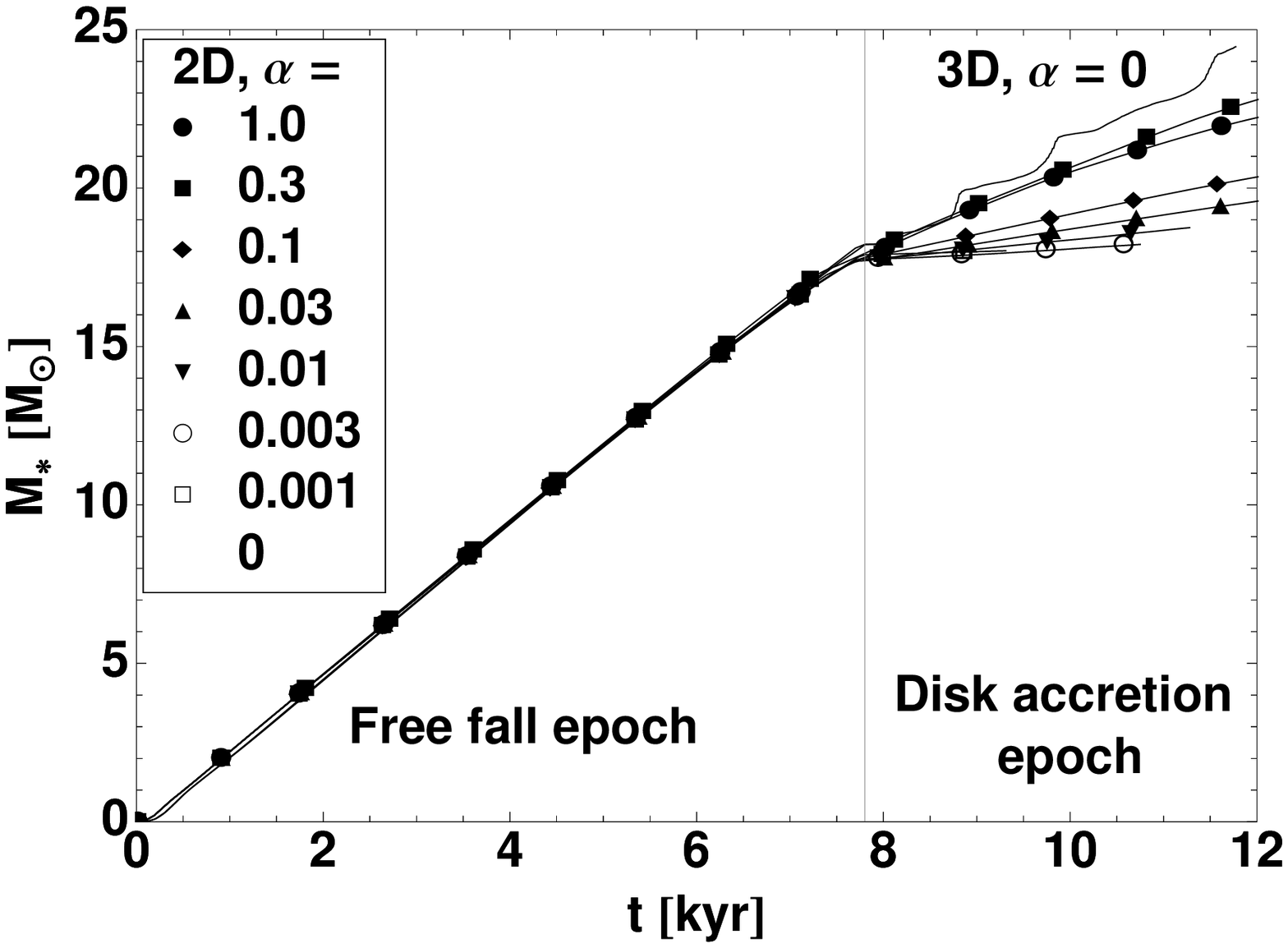}

\vspace{5mm}
\includegraphics[width=0.9\FigureWidth]{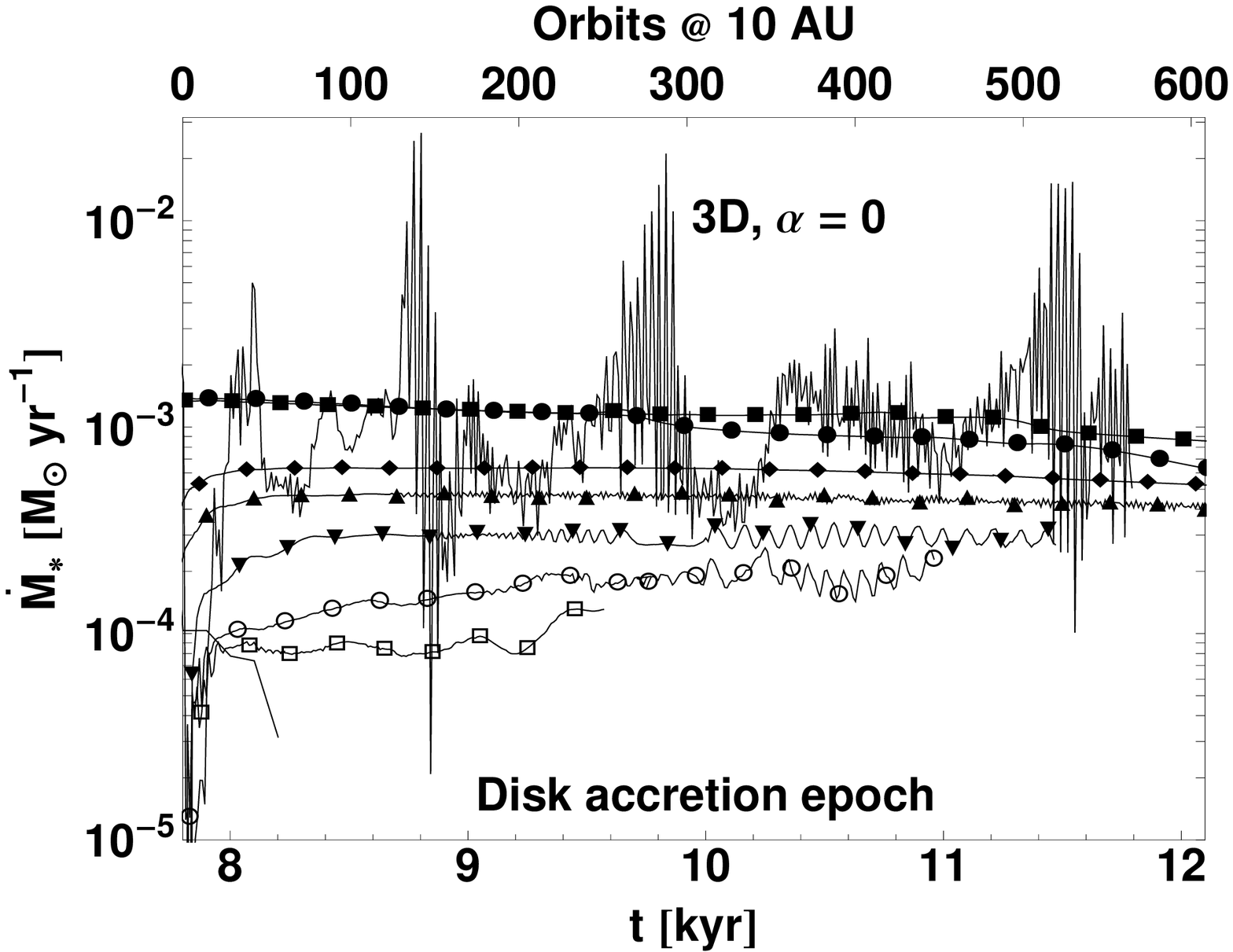}
\end{center}
\caption{
Stellar mass $M_*$ (upper panel) and accretion rate $\dot{M}_*$ (lower panel) as a function of time $t$ 
\vONE{
or number of inner orbits
}
during the collapse of a 120\Msol pre-stellar core.
The angular momentum transport in the two-dimensional axially symmetric run (lines with symbols) is driven by shear viscosity, calculated via an $\alpha$-parameterization prescription.
The angular momentum transport in the three-dimensional run (solid line) is driven by gravitational torques in the self-gravitating disk, see~Fig.~\ref{fig:Torques} for a visualization of the non-axisymmetric disk structure.
\vONE{
To visualize in particular the details during the disk accretion epoch, the lower panel displays the section from 7.8~kyr up to 12.1~kyr; as in the 60\Msol case, the accretion rate during the free fall epoch ($t < 7.8$~kyr) is constant.
}
}
\label{fig:3D}
\end{figure*}

To achieve a more detailed picture of massive accretion disks and to cross-check the assumptions made by applying an axially symmetric viscous disk model, we expand the two-dimensional setup into three dimensions now.
In the three-dimensional run no shear viscosity term is applied ($\alpha = 0$).
Instead, the forming massive accretion disk will self-consistently evolve non-axisymmetric structures yielding gravitational torques, which efficiently transport angular momentum to outer disk radii, allowing further accretion onto the new born star.
We perform the 3D simulation with an initial core mass of 120 $\mbox{M}_\odot$.
A comparison of the resulting stellar mass growth and the accretion history for the 3D case as well as several 2D cases with and without viscosity is presented in Fig.~\ref{fig:3D}.
\vONE{
Due to the high computational costs of the radiative 3D hydrodynamics simulations from the pre-stellar core collapse scale down to the 1~AU scale of the accretion disk formation and evolution, we are currently not able to report on the long-term evolution of the pre-stellar core collapse. 
The 11.8~kyr of collapse evolution simulated correspond to 0.25~$t_\mathrm{ff}$ of the initial pre-stellar core.
For comparison, in our previous 2D collapse simulations in \citet{Kuiper:2010p17191}, also the case $\alpha=0.3$ herein, we followed the collapse over 10~$t_\mathrm{ff}$ until no mass is left in the computational domain; it is either accreted onto the central star or expelled over the outer boundary at 0.1~pc by radiative forces.
\newline
Here, we will focus on the formation of the disk and the evolution of the main disk accretion phase, which is well resolved in the 3D simulation as well.
The disk accretion epoch in the 120\Msol run starts roughly at 7.8~kyr and we follow the evolution up to 11.8~kyr.
These 4~kyr correspond to 566 inner disk orbits (at 10~AU) or 28 outer disk orbits (on average roughly at 75~AU) respectively, based on a mean stellar mass of $M_* (7.8-11.8 \mbox{ kyr}) \approx 20\Msol$.
}

Compared to the axially symmetric simulations of the viscous accretion disk, the angular momentum transport by gravitational torques \vONE{due to the 3D non-axially symmetric disk morphology} results in a much more time dependent accretion history (lower panel of Fig.~\ref{fig:3D}), rather successive accretion events than a smooth accretion flow.
On the other hand, the integral over the accretion history $\int_0^t \dot{M}(t) \mbox{ d}t$, the actual stellar mass $M_*(t)$, remains qualitatively quite similar so far and finally yields a slightly higher total stellar mass (upper panel of Fig.~\ref{fig:3D}).

\clearpage
\section{Discussion}
\label{sect:Discussion}
\subsection{From the free fall phase to the disk accretion epoch}
A characteristic feature of the accretion histories, see lower panel of Fig.~\ref{fig:2D}, is a sharp drop down at roughly $t=8$~kyr.
This point in time marks the transition between the initial 'free fall' phase and the disk formation epoch afterwards.
During the free fall phase, the gas dynamics are dominated by the gravitational drag towards the core center.
During the course of the collapse, high angular momentum gas from initially larger radii reach the inner core.
At a specific point in time, the resulting centrifugal force (with slight support by thermal pressure) completely balances the gravity at the inner computational boundary (in the first grid cell adjacent to the central sink cell).
Hence the accretion rate drops down.

From now on, the follow-up mass spiraling inwards builds up a circumstellar disk.
In other words: The so-called centrifugal radius is increasing in time and therefore a circumstellar disk grows during the collapse from the inside outwards.
Further accretion through these disks has to be attended by an efficient angular momentum transport.

\subsection{Viscous disks (2D)}
Our simulation series in axial symmetry with varying values of the $\alpha$-shear-viscosity can be separated into two distinguishable regimes.
The runs with low viscosity ($\alpha \lesssim 0.05$) yield the formation of rings instead of the formation of an extended accretion disk.
As stated by \citet{Yorke:1995p1426}, these rings would break up in corresponding three-dimensional studies, resulting in angular momentum transport by tidal torques.
Runs with higher $\alpha$-viscosity yield the formation of long-living accretion disks.
Compared to analytical estimates on the $\alpha$-values of massive accretion disks considering typical high mass accretion rates in standard accretion disk models (including the thin disk approximation), presented in \citet{Vaidya:2009p12873}, 
\vONE{
our dynamical study allows also the formation of massive accretion disks with slightly larger $\alpha$-values.
}

\begin{figure*}[p]
\begin{center}
\includegraphics[width=\FigureWidth]{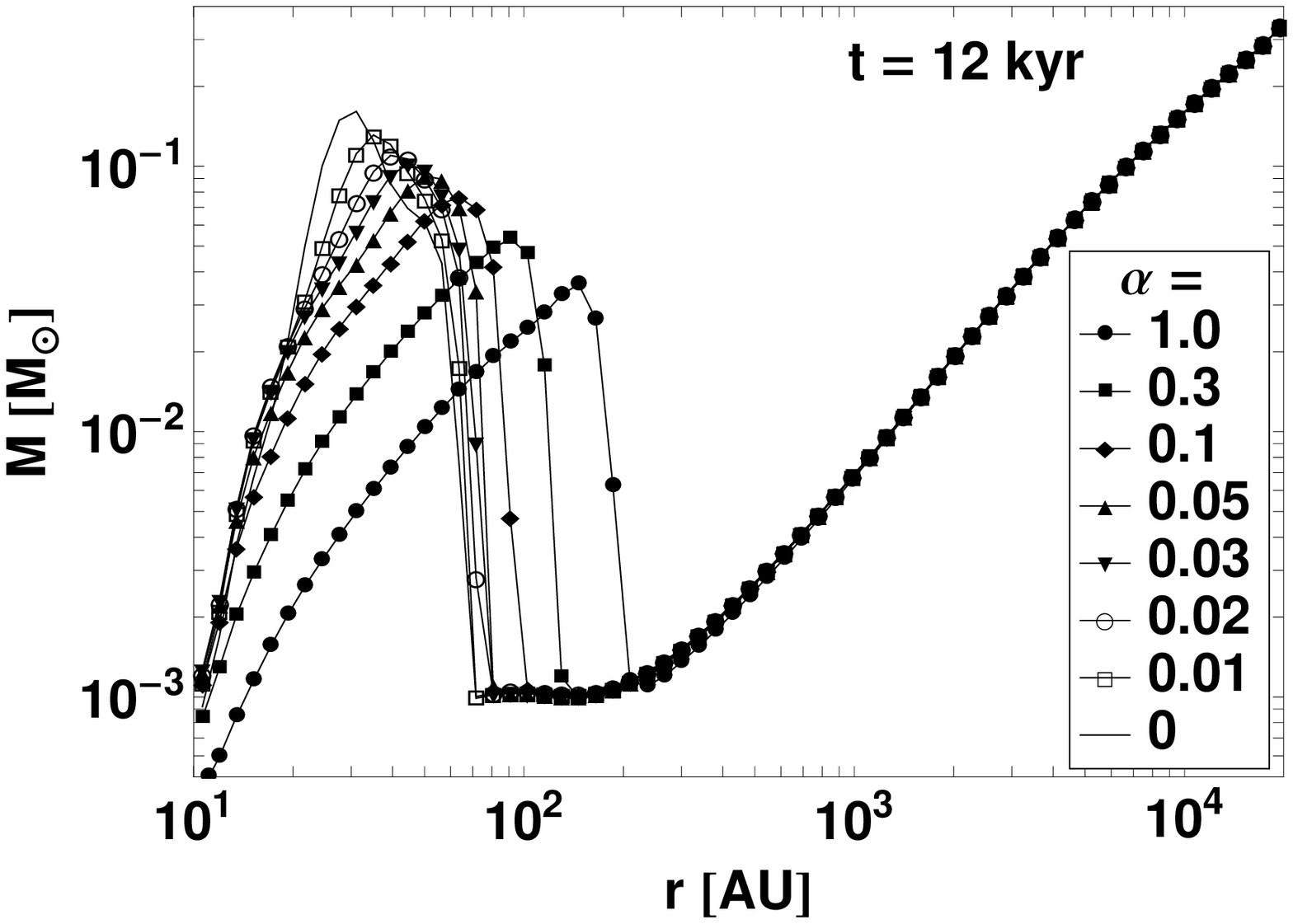}

\vspace{5mm}
\includegraphics[width=\FigureWidth]{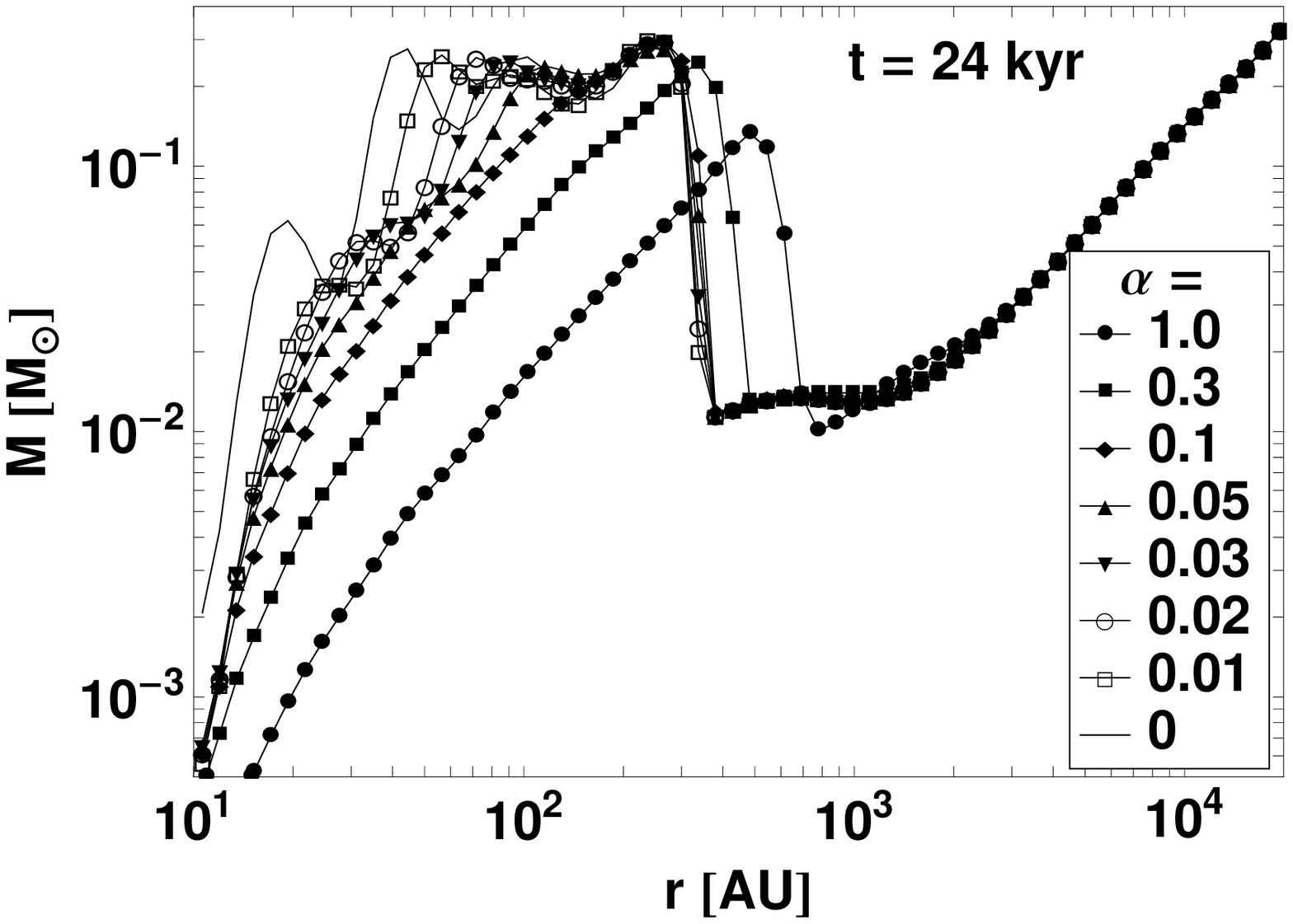}
\end{center}
\caption{
Gas mass $M$
in the domain's midplane 
(from midplane up to the first pressure scale height $H = 0.1 R$)
as a function of the radius $r$ 
for varying values of the $\alpha$-parameter
for two different points in time.
}
\label{fig:2D_DiskMass}
\end{figure*}

\begin{figure*}[htbp]
\begin{center}
\includegraphics[width=\FigureWidth]{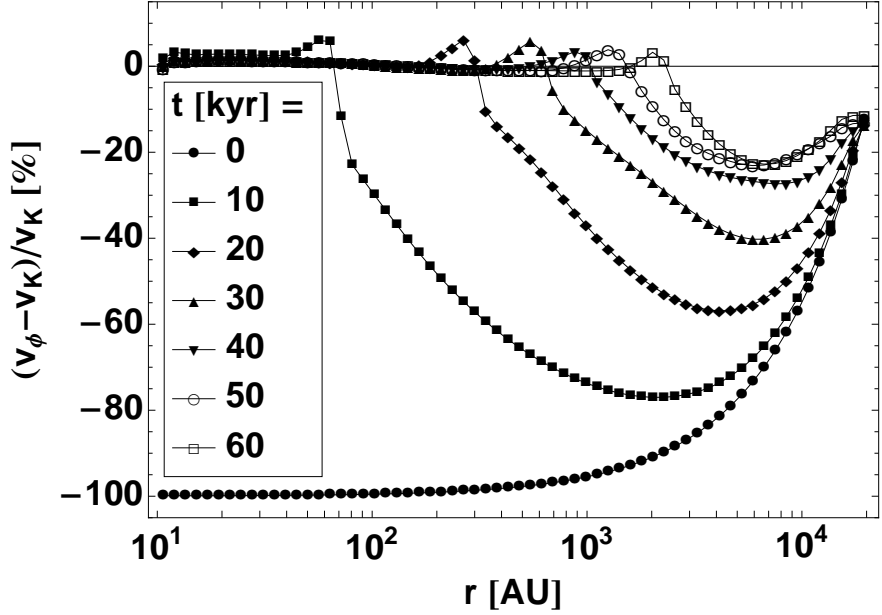}
\end{center}
\caption{
Deviation of the azimuthal velocity $v_\phi$ from Keplerian motion in the domain's midplane for the case of $M_\mathrm{core}=60\Msol$ and $\alpha=0.3$ for different points in time.
The Keplerian velocity $v_\mathrm{K}$ is hereby calculated based on the stellar mass plus the enclosed mass $M(r)$ in the computational domain.
The inner accretion disk is characterized by Keplerian motion.
The outer edge of the growing accretion disk is characterized by a super-Keplerian peak due to the newly in-falling material, which is not in gravitational-centrifugal equilibrium yet.
The outer envelope is characterized by a sub-Keplerian profile towards the initial solid body rotation ($t=0$).
}
\label{fig:2D_Kepler}
\end{figure*}

\begin{figure*}[p]
\begin{center}
\includegraphics[width=\FigureWidth]{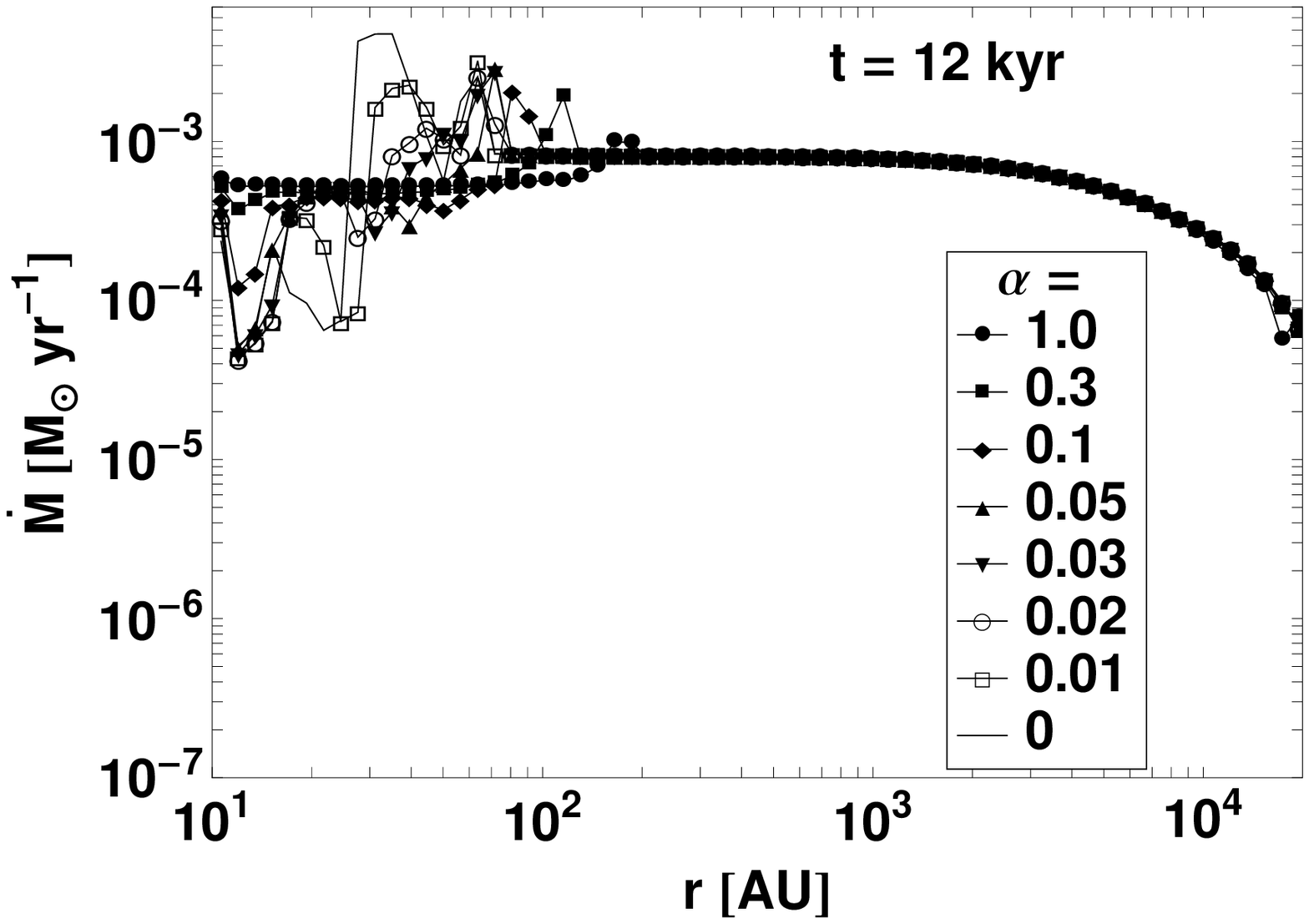}

\vspace{5mm}
\includegraphics[width=\FigureWidth]{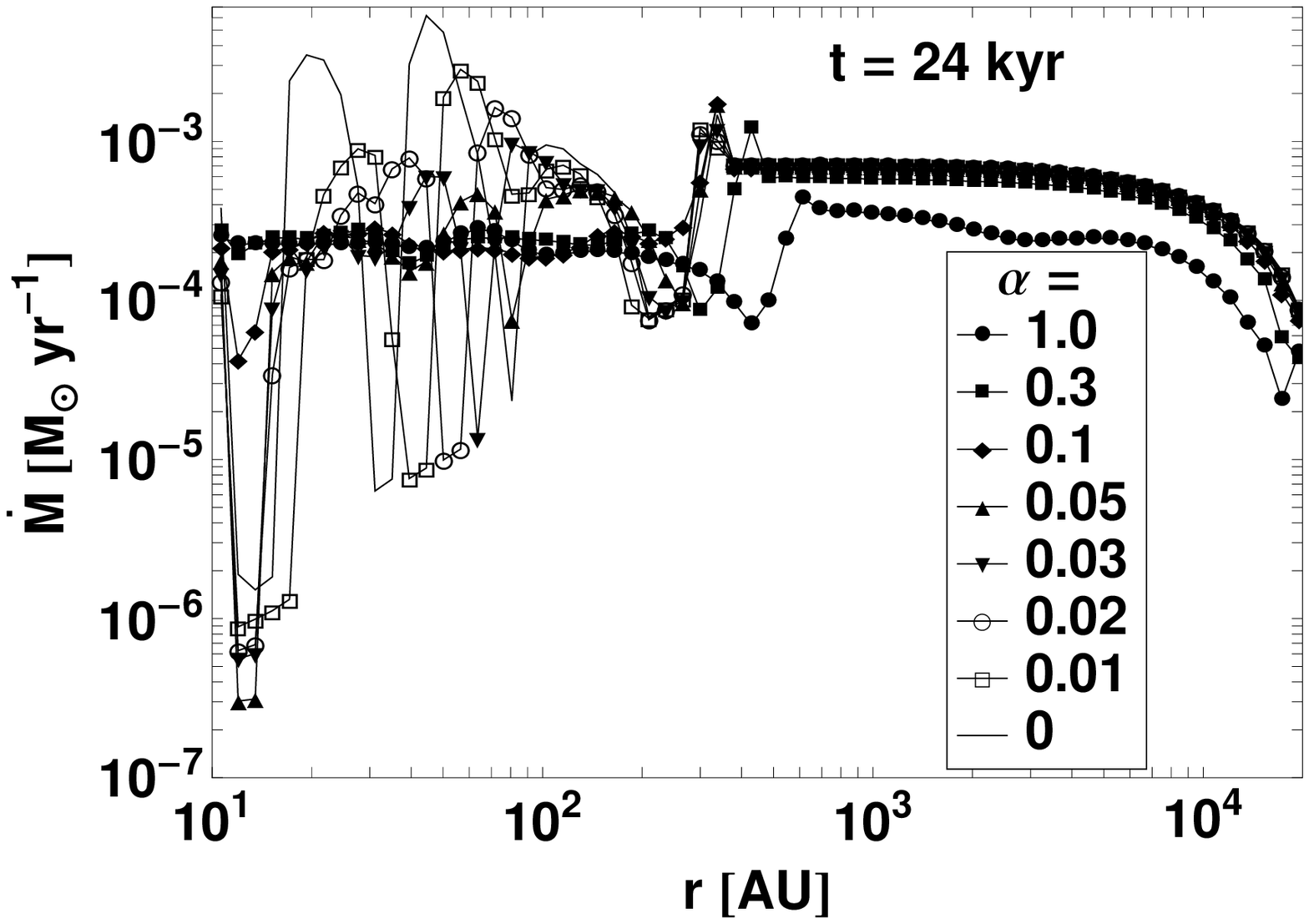}
\end{center}
\caption{
Accretion rate $\dot{M}$
integrated over the polar angle $\theta$
as a function of the radius $r$ 
for varying values of the $\alpha$-parameter
for two different points in time.
}
\label{fig:2D_AccretionRate}
\end{figure*}

During the long-term evolution of these viscous circumstellar disks, the accretion rate onto the central star decreases continuously.
The \vONE{long-term} runs
differ only about 16\% in the resulting stellar mass after $10^5$~yrs of evolution, 
\vONE{
although the $\alpha$-values of these runs vary up to a factor of 20.
After $10^5$~yrs the star has grown up to 19.0, 19.1, 17.8, and 16.0\Msol for $\alpha =$ 1.0, 0.3, 0.1, and 0.05 respectively.
}
If we infer from these simulations that the accretion rate is \vONE{rather} independent on $\alpha$, the well-known relation for the accretion rate in analytical steady state models
\begin{equation}
\dot{M}_* = 3 \pi \nu \Sigma 
\end{equation}
implies that the surface density $\Sigma$ is inversely proportional to the dynamical viscosity $\nu$, as it is expected from self-similar solutions for cases, in which the viscosity has a power-law dependence on the radius \citep[see e.g.][]{Hartmann:1998p14708}.
\vONE{
To backup this conclusion, Fig.~\ref{fig:2D_DiskMass} shows the mass in the midplane (up to the first pressure scale height $H = 0.1 R$ of the disk or $\Delta \theta = 5.625\degr$ respectively) as a function of the radius for varying values of the $\alpha$-parameter for two different points in time.
Due to the fact that $\alpha$ is only the normalization of the strength of the shear-viscosity, 
the slope of the in-falling envelope at radii larger than the disk radius,
where the shear force is negligible, 
is independent of the $\alpha$-parameter.
Going to smaller radii, we observe a characteristic steep increase in mass.
This accumulation of the in-falling mass in the midplane is simply a results of the conservation of angular momentum.
Each volume element at the initial radius $r$ and the polar angle $\theta$ can be related to its so-called centrifugal radius $R_\mathrm{cent}$, where gravity is completely compensated by the centrifugal force:
\begin{equation}
R_\mathrm{cent} = \frac{\Omega_\mathrm{0}^2 r^4 \sin(\theta)^2}{G~M(r)}
\end{equation}
with the initial angular velocity $\Omega_\mathrm{0}$ and the enclosed mass $M(r)$ inside the radius $r$.
This equilibrium of gravity and centrifugal force occurs at the inner computational boundary (at 10~AU) at roughly 7.8~kyr, afterwards the region of equilibrium increases radially.
If we define the location of the steep increase in mass as the outer disk radius $R_\mathrm{disk}$, we can infer a rough estimate on the growth rate of the circumstellar disk:
In the case of $\alpha=0.3$, the outer disk radius $R_\mathrm{disk}$ follows up to the first free fall time ($t_\mathrm{ff} = 67.6$~kyr) of the pre-stellar core roughly the linear relation
\begin{equation}
R_\mathrm{disk} \approx \frac{29 \mbox{ AU}}{1 \mbox{ kyr}} \left(t-t_\mathrm{onset}\right) + R_\mathrm{onset}
\end{equation} 
with the beginning of the disk formation at $t_\mathrm{onset} = 7.8$~kyr and $R_\mathrm{onset} = 10$~AU.
It has to be kept in mind that the outer radius $R_\mathrm{disk}$ of this accretion disk is defined via the steep increase in mass and the outer parts of this region $r \lesssim R_\mathrm{disk}$ are not necessarily in Keplerian rotation, see Fig.~\ref{fig:2D_Kepler}.
At the outer disk radius, the in-falling new disk material is not in gravitational-centrifugal equilibrium.
The material in motion falls down to a radius smaller than its centrifugal radius and therefore this region of disk growth is characterized by super-Keplerian motion.
For an observational example of super-Keplerian motion around young forming high-mass stars, see \citet{Beuther:2008p4374}.
At larger radii, the envelope - still feeding the accretion disk - shows a strong sub-Keplerian profile.
This outcome supports the speculation by \citet{Beuther:2009p12913}, who state that no massive accretion disk in Keplerian motion could be observed so far due to the current observational resolution limit of roughly 1000~AU.
But embedded in the currently observed large-scale flattened structures in non-Keplerian motion, they expect to find Keplerian massive accretion disks with radii smaller than 1000~AU.
The size of the disks simulated here expands in time.
\newline
In the sheared disk region ($r < R_\mathrm{disk}$) higher values of $\alpha$ indeed result in lower mass disks.
As discussed previously, for low viscosity runs ($\alpha \le 0.05$) the evolution leads to the formation of rings instead of a stable accretion disk.
\newline\indent
The accretion rates as a function of the radius are displayed in Fig.~\ref{fig:2D_AccretionRate}.
In the low-$\alpha$-simulations with $\alpha \le 0.05$, in which rings have formed rather than accretion disks, the accretion rate varies strongly with the radius.
If the $\alpha$-viscosity is high enough to maintain the accretion flow, the accretion rates are indeed approximately constant throughout the whole disk region and also independent of the strength of the $\alpha$-viscosity.
}

In terms of the pre-stellar core collapse simulations presented here this means that the impact of the choice of the $\alpha$-value on the accretion rate is almost negligible (unless the $\alpha$-value is chosen high enough), but strongly affects the morphology of the disk.
\vONE{In the following section, we compare these axially symmetric $\alpha$-disk models with a}
self-consistent 
\vONE{scenario, in which} the angular momentum transport
is calculated from the disk evolution itself.

\clearpage
\subsection{Self-gravitating disks (3D)}
\begin{figure*}[p]
\centering
\parbox{0.5\FigureWidth}{
\includegraphics[width=0.5\FigureWidth]{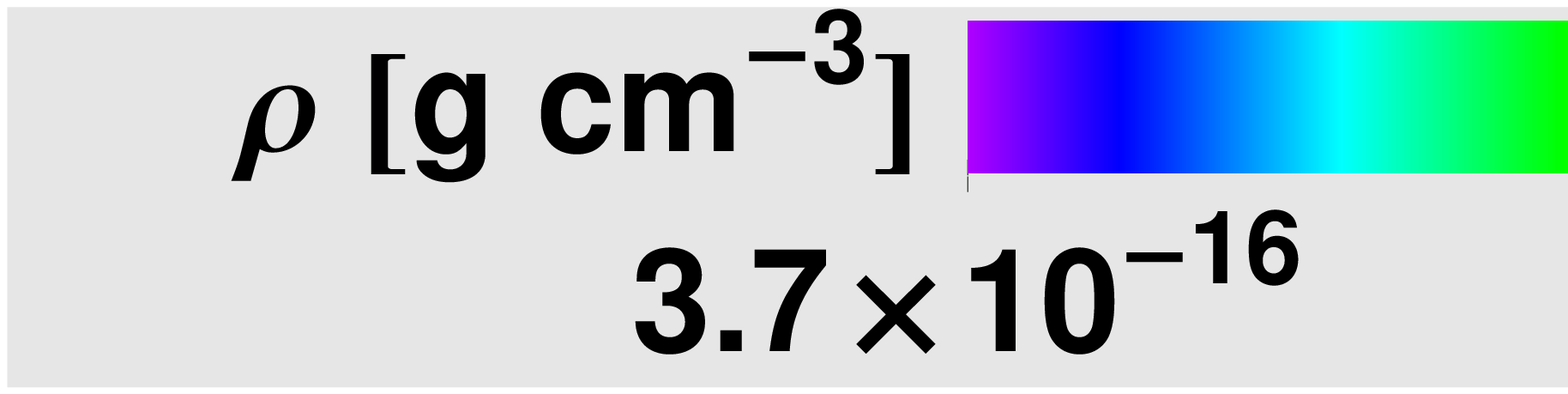}
\includegraphics[width=0.5\FigureWidth]{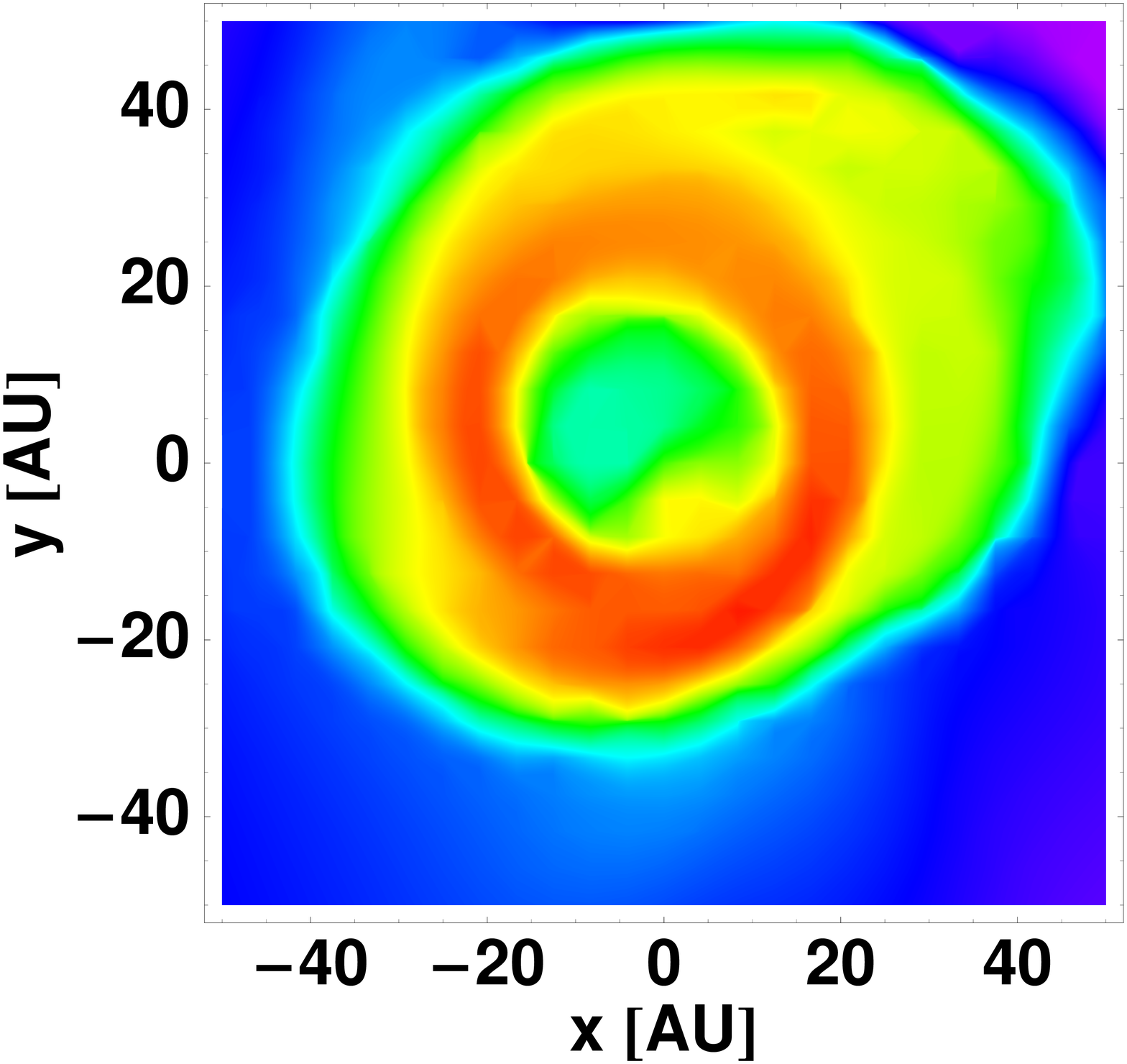}
}
\hspace{10mm}
\parbox{0.5\FigureWidth}{
\includegraphics[width=0.5\FigureWidth]{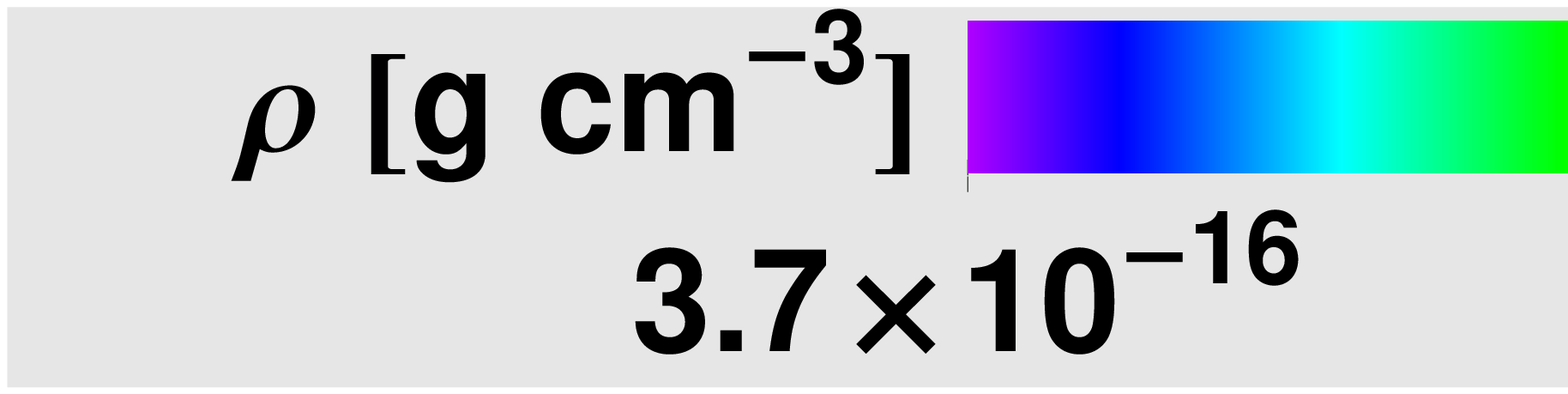}
\includegraphics[width=0.5\FigureWidth]{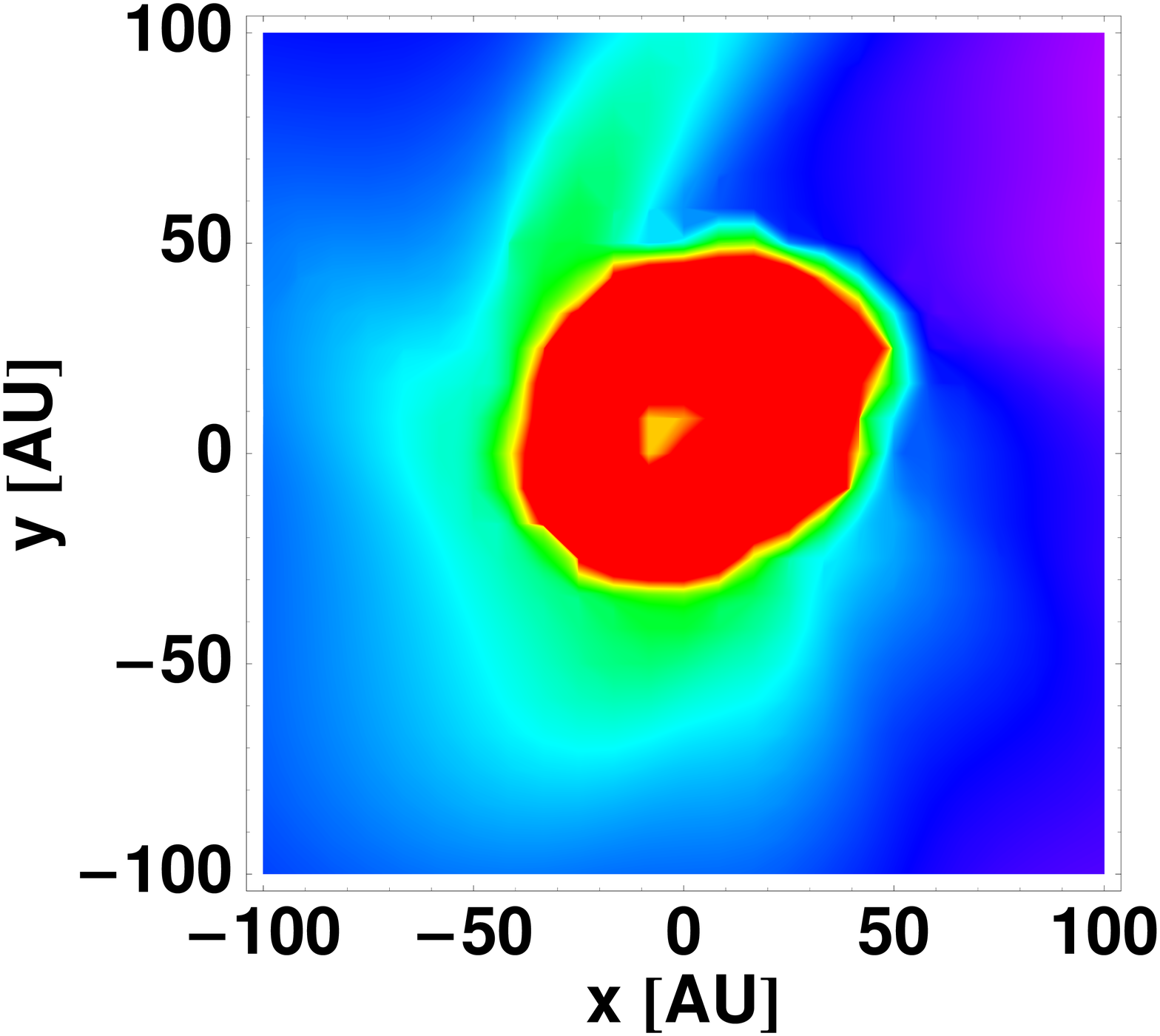}
}
\vspace{10mm}

\parbox{0.5\FigureWidth}{
\includegraphics[width=0.5\FigureWidth]{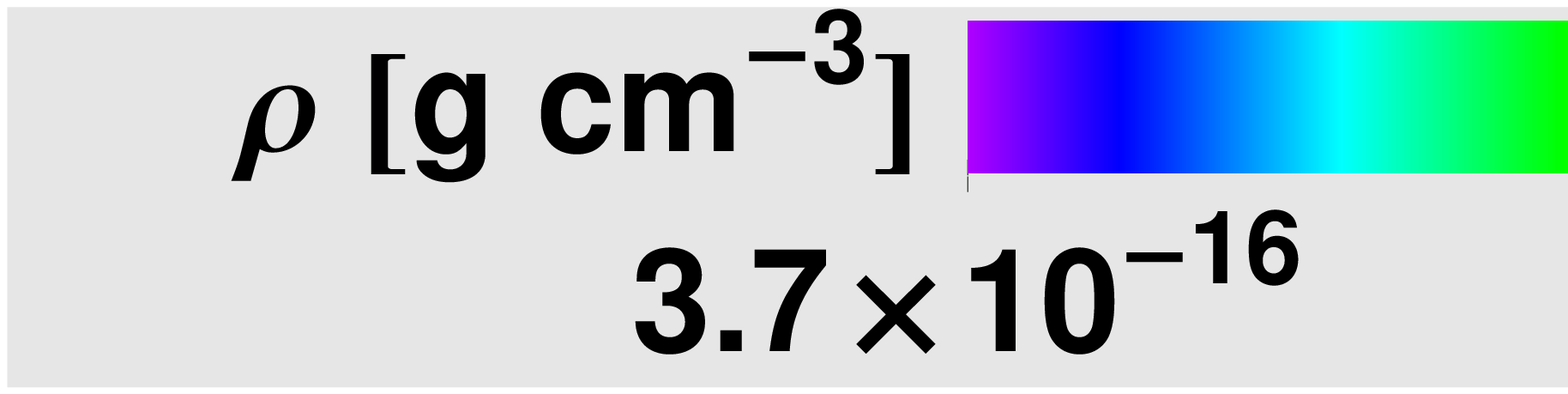}
\includegraphics[width=0.5\FigureWidth]{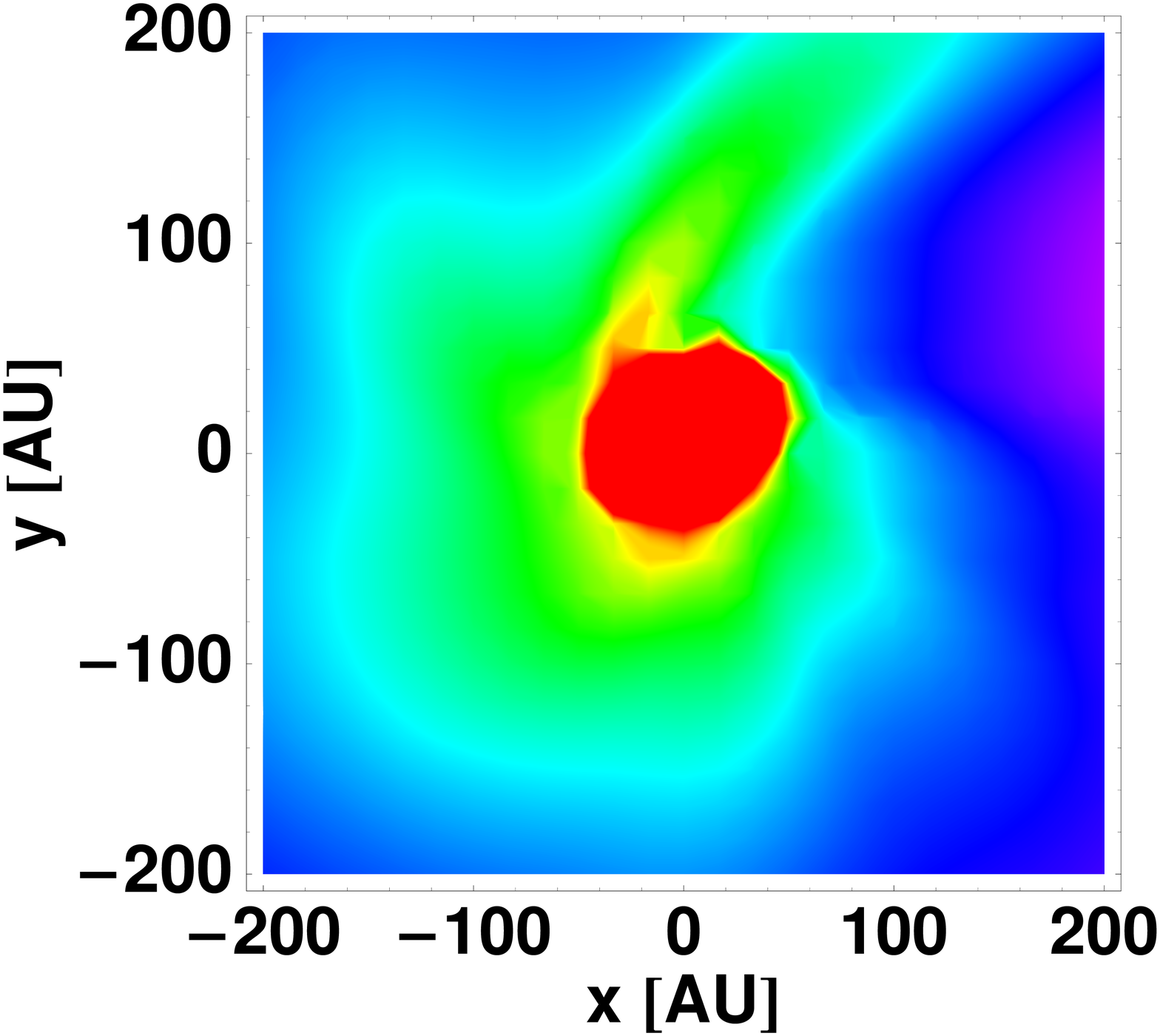}
}
\hspace{10mm}
\parbox{0.5\FigureWidth}{
\includegraphics[width=0.5\FigureWidth]{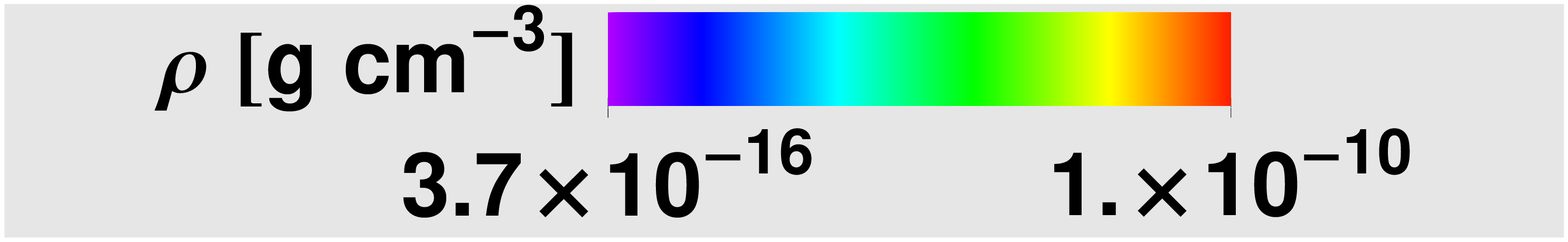}
\includegraphics[width=0.5\FigureWidth]{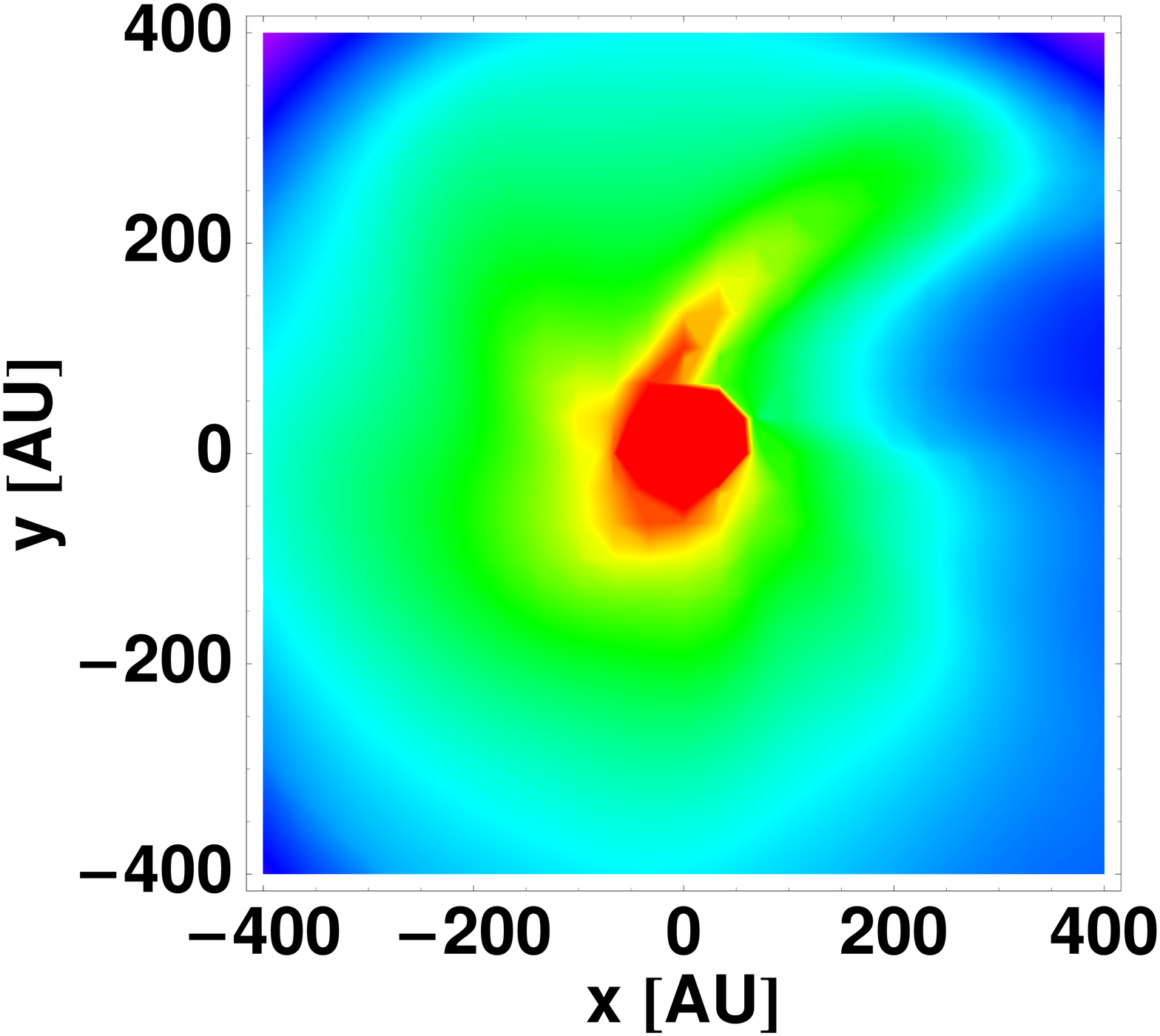}
}
\caption{
Face-on view of the high density gas in a slice through the $x$-$y$~midplane.
The length of the clipping domains goes from 50 AU x 50 AU up to 400 AU x 400 AU.
The data is taken from a snapshot ($t=10$~kyr) of the 120\Msol pre-stellar core collapse.
A linear interpolation was applied to the discretized data cube.
The density of the central sink cell with a radius of 10 AU is calculated by an extrapolation of the gas density at the boundary assuming a zero gradient in the density, so this region should be taken with care.
}
\label{fig:Torques}
\end{figure*}

Hence we started to expand our simulations into three dimensions.
These simulations allow to compute the angular momentum transport by developing gravitational torques in the self-gravitating disk consistently with the formation and evolution of the accretion disk.
A visualization of such a non-axisymmetric density structure yielding a gravitational torque is presented for the case of a 120\Msol collapse in Fig.~\ref{fig:Torques}.
The figure shows a face-on view of a slice through the inner core region at $t=10$~kyr.

The similarity of the mean accretion rate \vONE{with the high-$\alpha$-disks} justifies the assumption made in the viscous disk simulations, in which the computation of the azimuthal structure of the accretion disks is replaced by an a priori $\alpha$-parameterization.
The $\alpha$-prescription comprises to a reasonable accuracy the overall angular momentum transport by evolving instabilities in the three-dimensional disk.

The origin of these instabilities lies in the self-gravity of the circumstellar disk as supported by the fact that the Toomre $Q$ value 
\begin{equation}
Q = \frac{\Omega c_\mathrm{s}}{2 \pi G \Sigma}
\end{equation}
approaches unity from above ($Q \rightarrow 1$ and $Q > 1$) at the unstable radius during the formation of the spiral pattern due to an enhancement of the surface density.
Fragmentation of the disk during the further evolution of this instability is likely to be suppressed by the heat as well as the fast rotation of the inner disk.
The Toomre $Q$ value in the unstable regime stays above unity so far ($1 < Q < 2$).

Modes with lower azimuthal wave number (especially $m=1$) are potentially susceptible for a growth by the so-called SLING instability ("{\bf S}timulation by the {\bf L}ong-range {\bf I}nteraction of {\bf N}ewtonian {\bf G}ravity"), which was first studied numerically in \citet{Adams:1989p14769} and analytically in \citet{Shu:1990p14813}.
Nonetheless, amplifications by the SLING instability require a disk to stellar mass ratio of $M_\mathrm{disk} / M_\mathrm{star} > 1/3$.
The SLING instability seems to be the major driver of amplification up to the point of binary formation if $M_\mathrm{disk} / M_\mathrm{star} \rightarrow 1$, which could be the case during the early onset of a pre-stellar core collapse before the central star dominates \citep{Adams:1992p14918}.
Note, that $m=1$ modes cause the center of mass to be unequal to the geometrical center of the disk, which is not taken into account in our simulations.
Anyway, in our simulations the instability occurs when the star has already grown up to 20\Msol so that amplification by the SLING instability seems to be unlikely.
Furthermore, the amplification by the SLING instability only grows slowly on timescales of a few outer rotation periods of the disk (with a minimum of one outer rotation period for $M_\mathrm{disk} / M_\mathrm{star} = 1$).
Such a large mass ratio is observationally detected when comparing the large scale (non-Keplerian) flattened structure with the proto-stellar mass \citep[e.g.][]{Cesaroni:1997p4541, Zhang:1998p4666, Beuther:2005p1793}.
Contrary to this slow growth, the so-called SWING amplification of modes with larger azimuthal wave numbers $m>1$, proposed first by \citet{Goldreich:1965p14960} and \citet{Toomre:1981p14967}, potentially will grow much faster.
Especially once axisymmetric disturbances due to self-gravity become prominent, modes with $m>1$ dominate \citep[cp.][]{Papaloizou:1991p14835}.

\begin{figure*}[p]
\begin{center}
\includegraphics[width=\FigureWidth]{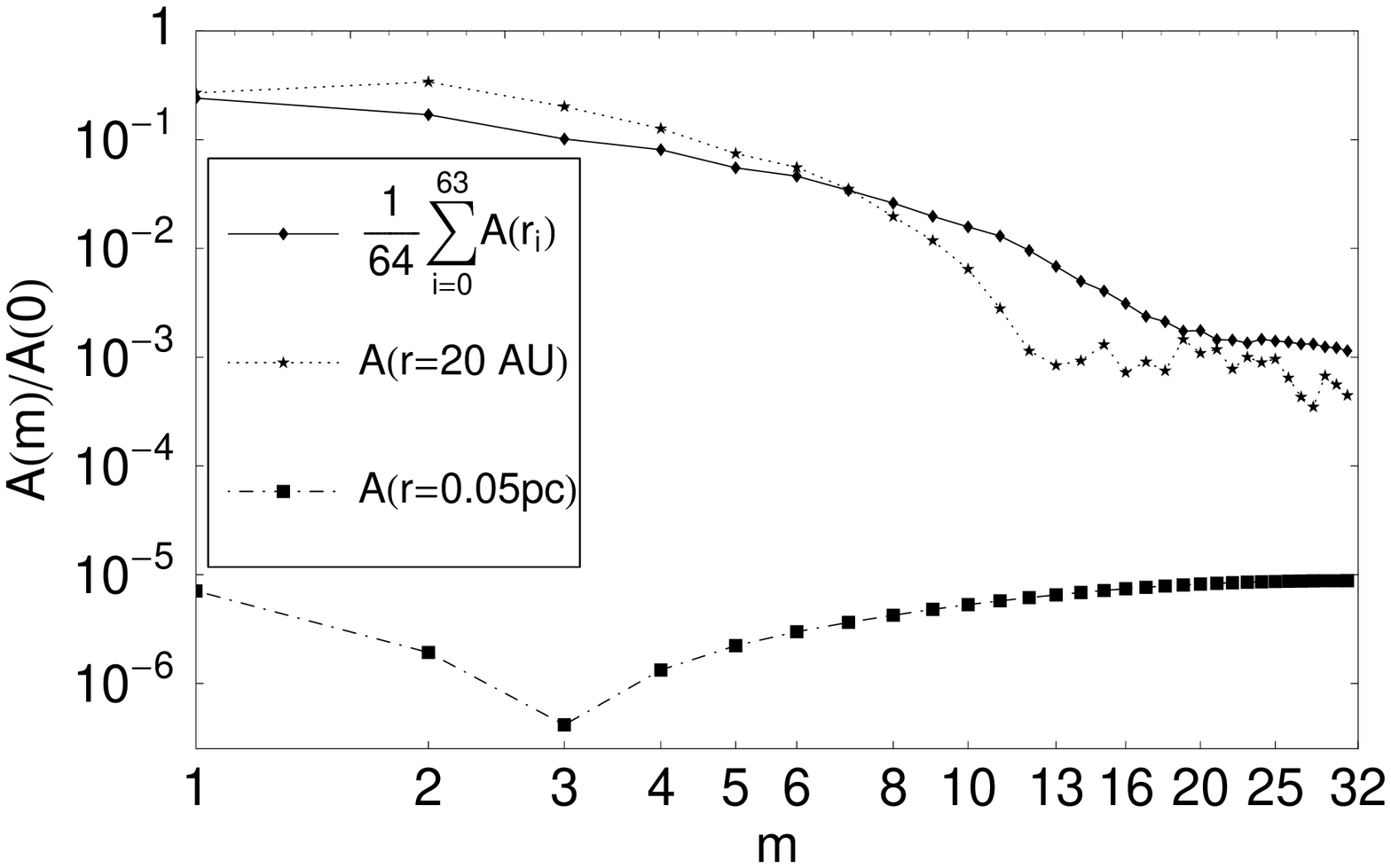}
\includegraphics[width=\FigureWidth]{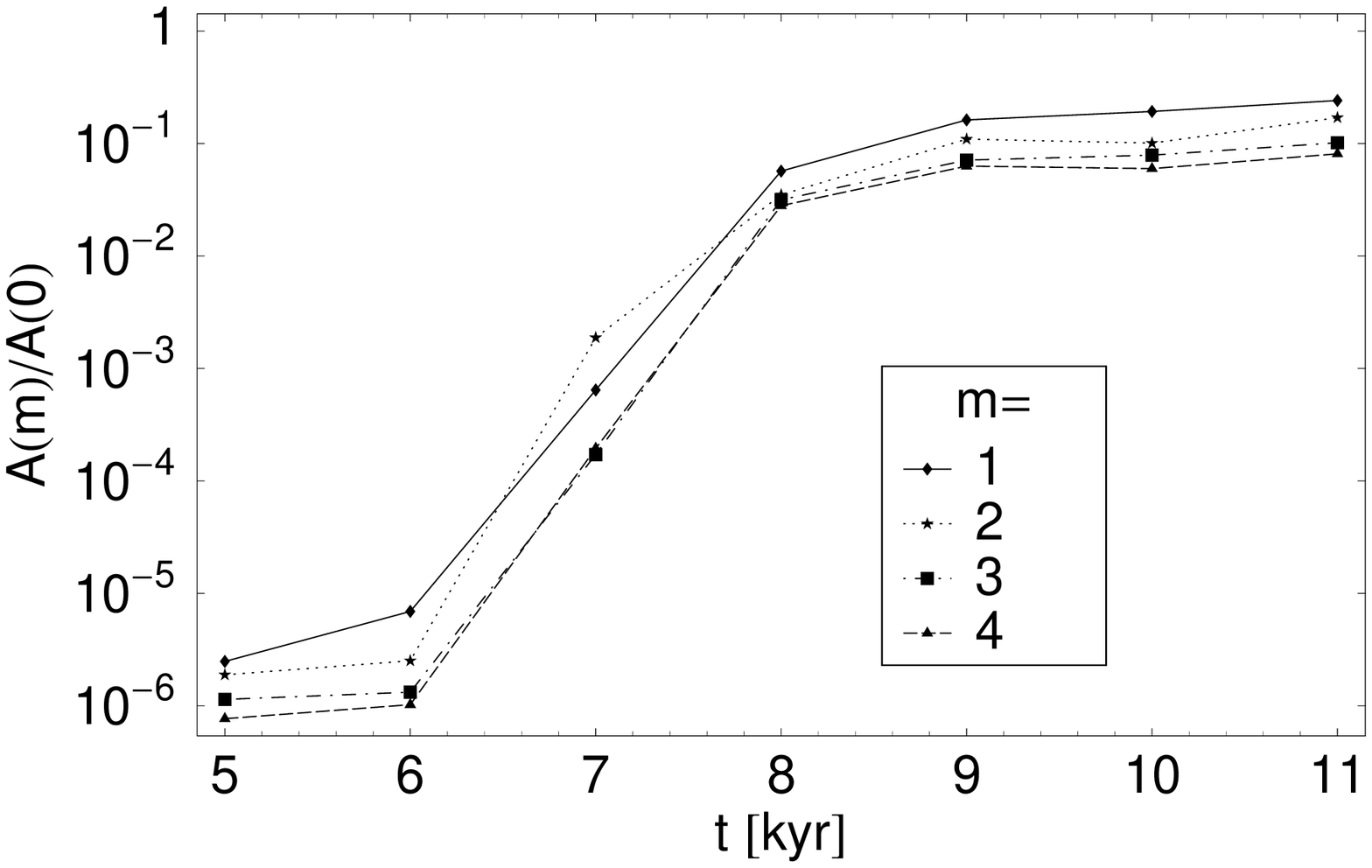}
\end{center}
\caption{
Fourier analysis of the non-axisymmetric accretion disk formed during the collapse of a 120\Msol pre-stellar core.
The upper panel displays the strength of the modes $m$ at a specific point in time $t=11$~kyr. 
The different lines clarify the dependency on the radius $r$ and denote the modes at the inner disk rim (dotted line with stars), at the outer core region (dashed dotted line with squares) as well as the mean amplitude of the total sum over all radii (solid line with diamonds).
The lower panel shows the evolution of the mean amplitude with time for the four highest modes $m=$ 1, 2, 3, and 4.
The amplitudes in both panels are normalized to the $m=0$ mode.
}
\label{fig:Modes}
\end{figure*}

The Fourier analysis of the non-axisymmetric disk formed in the 120\Msol collapse is displayed in Fig.~\ref{fig:Modes}.
The figure shows the amplitudes of the modes for different radii and the mean amplitude of the sum over all radii (upper panel) as well as the evolution of the mean amplitude with time (lower panel).
The upper panel documents that the $m=2$ mode has a higher amplitude than the $m=1$ mode at the inner rim of the disk
\vONE{($(A(m=2)-A(m=1))/A(m=1) = 26\%$)}. 
Elsewhere $m=1$ is slightly higher.
The evolution of the mean amplitude (normalized sum over all radii) with time, shown in the lower panel, demonstrates that during the onset of the gravitational instability the $m=2$ mode grows faster than the others.
The mean growth rate is roughly independent on the azimuthal wave number $m$.

Further well-founded conclusions on the evolution (the growth of unstable modes with different azimuthal wave numbers) and fate (potential fragmentation) of the spiral patterns require to study the disk's evolution in the three-dimensional run for a much longer period (potentially up to a factor of 40 for a complete coverage of the stellar accretion phase) and to analyze a variety of initial conditions.
Moreover, a study of the potential fragmentation of the three-dimensional disk structure is currently inhibited in these simulations due to the fact that after the onset of spiral arm formation the Truelove criterion is violated.
The Jeans number $J=\Delta x / \lambda_\mathrm{J}$ in the high-density regions approaches unity, i.e.~the Jeans length $\lambda_\mathrm{J} = \sqrt{\pi c_\mathrm{s}^2 / G \rho}$ is only resolved by a single grid cell with the resolution $\Delta x$.
\citet{Truelove:1997p10742} recommended a Jeans number of $J \le 0.25$, i.e.~to resolve the Jeans length by at least four grid cells.
Hence, the best and straight forward improvement would be to go to an even higher resolution of the forming circumstellar disk.
An alternative approach, used e.g.~in \citet{Krumholz:2007p1380, Krumholz:2009p10975}, would be to use so-called sink particles when the Truelove criterion is violated. 
In the sink particle method, the Truelove criterion implies a maximum density, which is considered in the gas dynamics (regions with higher density are represented by the particles).
The method of sink particles should be justified, if it is guaranteed that the gas accumulated in the sink would be gravitationally bound anyway. 
On the other hand, this study focuses on the resulting accretion history of the forming massive star, which initial phase shown here completely relies on the well resolved large scale structure of the non-axisymmetric spiral arms.
Fragmentation on larger scales \vONE{potentially} limits the mass reservoir of the forming star in the center \citep{Krumholz:2009p10975, Peters:2010p12919, Peters:2010p17050}, although \citet{Peters:2010p16326} recently have shown that a large scale magnetic field will offer locally support against gravitational collapse reducing secondary fragmentation.

\vONE{
The results of our disk accretion simulations match the previous models of viscous and self-gravitating disks by \citet{Vorobyov:2009p16255}, proposing viscous-dominated disks for $\alpha > 10^{-2}$ and very high disk accretion rates for $\alpha > 10^{-1}$.
The episodic accretion events in our self-gravitating three-dimensional simulations and the non-episodic accretion of our two-dimensional viscous disks are in good agreement with the results by \citet{Vorobyov:2010p16256}, who state that ``the (additional) effect of a generic $\alpha$-type viscosity acts to reduce burst frequency and accretion variability, and is likely to not be viable for values of $\alpha$ significantly greater than 0.01''.
}
For a comparison of these outcomes with previous semi-analytic predictions including disk fragmentation, please see also \citet{Kratter:2006p11356, Kratter:2008p17135}.

A direct comparison of the resulting accretion histories to observations suffer from the fact that currently the stellar accretion rate cannot be observed in the case of high-mass star formation due to the high extinction of the proto-stellar environment.
Observations with reference to low-mass star formation clearly favor the episodic accretion events shown in our three-dimensional simulation, see e.g.~\citet{Enoch:2009p16308, vanBoekel:2010p14739} for most recent examples.
Previous three-dimensional simulations by \citet{Krumholz:2007p1380, Krumholz:2009p10975} fully support this outcome, too.

\clearpage
\section{On the stability of radiation pressure driven outflows}
\label{sect:RRTI}
Although this publication highly focuses on the angular momentum transport and accretion flow through the circumstellar disk, the three-dimensional simulation reveal more interesting physical processes.

\subsection{The issue}
Background of this section is that our simulation series in axial symmetry presented in \citet{Kuiper:2010p17191} have shown the launch of a stable radiation pressure driven outflow in the bipolar direction.
As a result, the radiation pressure removes a substantial fraction (more than $40\%$ during the stellar disk accretion phase) of the initial mass of the pre-stellar core from the star-disk system, reducing the star formation efficiency of the massive star forming core.

\begin{figure*}[p]
\centering
\includegraphics[width=\FigureWidth]{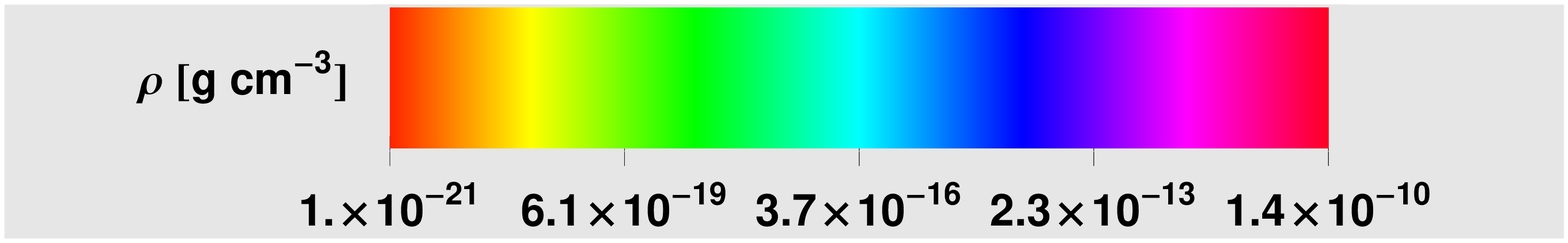}

\includegraphics[width=\FigureWidth]{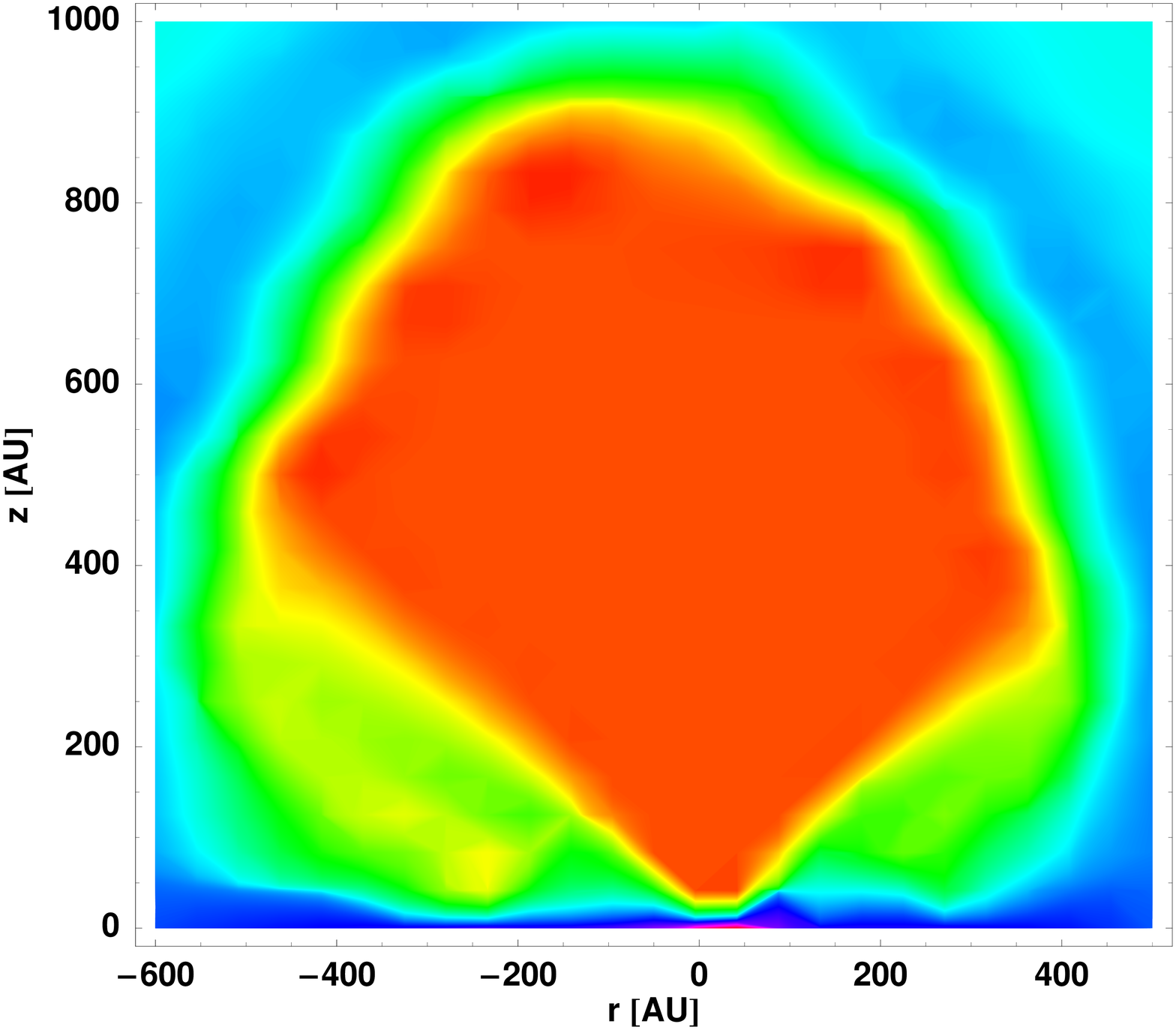}
\caption{
Edge-on views of the low density gas in a slice through the $x$-$z$~plane.
The data is taken from a snapshot ($t=10$~kyr) of the 120\Msol pre-stellar core collapse.
A third order interpolation was applied to the discretized data cube.
}
\label{fig:Outflow}
\end{figure*}

In the simulations by \citet{Krumholz:2009p10975}, these outflows are affected by the so-called '3D radiative Rayleigh-Taylor instability'. 
In our three-dimensional simulation presented here, the strong radiation pressure in the bipolar direction stops the in-fall and reverts the mass flow into a radiation pressure driven outflow with velocities of a few $100 \mbox{ km s}^{-1}$.
Fig.~\ref{fig:Outflow} visualizes a slice through the three-dimensional density structure of such an outflow, launched during the collapse of a 120\Msol pre-stellar core.
In contrast to \citet{Krumholz:2009p10975}, the outflow in our three-dimensional simulation remains stable despite the strong non-axially symmetric features developing in the outflow region so far.

\subsection{Possible explanations}
The reason for the disagreement on the stability of the radiation pressure driven outflow can be manifold.
Although we cannot guarantee the completeness of the following listing, we try our best to objectively overview the most relevant explanations here:
\begin{enumerate}
\item \vONE{We use an initially steeper density profile ($\rho \propto r^{-2}$) than \citet{Krumholz:2009p10975} ($\rho \propto r^{-1.5}$). I.e., at the point in time, when the radiation pressure driven outflow is launched, there is more mass in the bipolar direction, which is swept up at the border of the developing cavity, in the configuration by \citet{Krumholz:2009p10975}.}
\item The instability potentially requires non-axial symmetric modes (3D). 
Therefore our 2D runs, including the whole stellar accretion phase, show a stable and long-living outflow. 
In our 3D run, the outflow is stable during the launching and the first growth phase, but the instability will occur at a later epoch, e.g.~the instability in \citet{Krumholz:2009p10975} occurs when the outflow extents up to slightly more than $r \sim 1000$~AU and the outflows in our runs have just propagated up to $r \le 1000$~AU so far.
\item Our simulations account for frequency dependent stellar irradiation instead of the frequency averaged (gray) approximation used in \citet{Krumholz:2009p10975}. 
In case of gray radiation transport, the stellar irradiation is absorbed in a thin layer (the 'radiation bubble wall' in \citet{Krumholz:2009p10975}) independent of its natural frequency.
In case of frequency dependent irradiation, the IR part of the stellar spectrum is absorbed behind the UV absorption layer.
This potentially results in a pre-acceleration of the layers on top of the actual cavity.
The sharp wall as seen in \citet{Krumholz:2009p10975} changes to a broader transition region, if the frequency dependence is considered.
\item Additionally to the frequency dependence, our simulations use a more accurate treatment (ray-tracing) of the `first absorption' of the stellar irradiation in the cavity wall, and the instability is an artifact of the flux-limited diffusion approximation.
The diffusion approximation, the assumption that photons behave like a diffuse gas, is only valid in an optically thick medium.
The flux limiter guarantees also for an optically thin medium that the radiation velocity is always less than or maximal equal to the speed of light, but does not represent the correct motion of photons: In the optically thin cavity, photons will propagate on straight lines impinging on the cavity wall, instead of a diffuse expansion such as a gaseous bubble. Exactly in the cavity wall, at $\tau \sim 1$, the flux-limited diffusion approximation breaks down and yields the wrong radiative flux.
\item The disagreement is due to the different numerical resolution of the diverse grid structures:
Under the assumption that the refinement criteria of the AMR grid guarantees the usage of the highest resolution level at the cavity walls ($\Delta x=10$~AU in the case of \citet{Krumholz:2009p10975}), our fixed and radially stretched grid in spherical coordinates has the same resolution of $\Delta x \sim 10$~AU roughly at a radius of $r=100$~AU.
In the inner core region ($r<100$~AU) the resolution of our fixed grid increases gradually up to $\Delta x \sim 1$~AU.
In the outer core region ($r>100$~AU) the resolution of our fixed grid decreases gradually down to $\Delta x \sim 2000$~AU.
Significant statements on the resolution needed require more resolution tests with both grid structures.
Furthermore, for a comparison of the different resolution, the following item has to be considered, too.
\item We use a grid in spherical coordinates, while \citet{Krumholz:2009p10975} use Cartesian.
Both dominant forces in the bipolar region (gravity and radiation pressure) are in first order aligned with the radial coordinate axis, i.e.~they are better represented in a spherical coordinate system.
This difference in the coordinate system gets unimportant by the time the resolution needed is known and used. 
But at least this points out that a higher resolution is needed when using Cartesian coordinates to represent the radially acting forces sufficiently (e.g.~analogously much higher resolution is needed in Cartesian coordinates to represent a circular orbit than in spherical coordinates, where the orbital motion is aligned with the azimuthal axis).
\end{enumerate}

\subsection{Conclusion}
The simulations by \citet{Krumholz:2009p10975} and ourselves \citep[][and the simulations herein]{Kuiper:2010p17191} differ in a large variety of numerical methods, initial conditions, and radiative physics included, which could be responsible for the disagreement on the stability of the outflow region.
We suppose that the underlying physical difference is the fact that the expansion of the radiation bubble is stopped in the simulations by \citet{Krumholz:2009p10975}.
I.e., the gravitational force is roughly as high as the radiation pressure in this situation (during the launch of the bubble the radiation pressure has to be higher).
Then the marginal stable bubble wall is subject to a radiative Rayleigh-Taylor instability.
But, in our opinion, this kind of instability would work in two dimensions as well and will not require non-axially symmetric modes to occur.
In our simulations, the central star still accretes mass after the launch of the outflow, yielding a continuous growth of its luminosity.
The radiation pressure remains stronger than gravity and the expansion of the optically thin cavity region is not stopped.
In this way, the radiation pressure expels more than half of the initial core mass from the star-disk system and therefore efficiently reduces the star formation efficiency.

To conclude: The details of the underlying physics of the `3D radiative Rayleigh-Taylor instability' are quite unknown so far and deserve further investigation.
A well-founded answer to the stability will have an impact on the more general question of the importance of feedback processes on the star formation efficiency.
Anyway, the final stellar masses of our axially symmetric simulations, including a stable long-living bipolar outflow, evidently support the  statement that instabilities in the outflow regions are not required to form the most massive stars.

\clearpage
\section{Summary}
\label{sect:Summary}
We performed axially symmetric and three-dimensional high-resolution radiation hydrodynamics simulations of monolithic pre-stellar core collapses towards the formation and evolution of massive accretion disks including frequency dependent radiative feedback.
The vicinity of the forming star could be resolved down to 1.27~AU.
We directly compared the history of accretion rates driven either by gravitational torques due to the self-gravity of the massive circumstellar disk or by an $\alpha$-shear-viscosity prescription.
The influence of the choice of the $\alpha$-value on the resulting accretion history was determined in a  simulation series.

We found that the accretion rates and therefore also the final stellar masses depend only marginally on the strength of the $\alpha$-viscosity as long as a sufficient transport of angular momentum is guaranteed to avoid the formation of ring instabilities.
\vONE{
Instead, the choice of the strength of the $\alpha$-shear-viscosity determines the morphology of the forming circumstellar disk in such a way that higher $\alpha$-values lead to lower surface densities.
}
Contrary to the continuous accretion during the viscous disk evolution,
the three-dimensional simulation results in time dependent episodically accretion events, as already observed in the case of low-mass star formation.
Nevertheless, the mean accretion rate is very similar to the $\alpha$-viscosity approach of the axially symmetric simulations (even a slightly higher mean accretion rate can be obtained by the developing gravitational torques) and therefore fully justifies the assumptions made in the two-dimensional studies.

Our simulations show the launch of radiation pressure driven outflows, which as long as we can proceed with the three-dimensional run maintain their stability in both axial symmetric and non-axial symmetric stellar environments.

Finally, the angular momentum transport via gravitational torques is shown to be sufficiently strong to allow for the high accretion rates needed during the formation of the most massive stars.

\acknowledgments
This research has been supported by the International Max-Planck Research School for Astronomy and Cosmic Physics at the University of Heidelberg (IMPRS-HD).
H.~Klahr has been supported in part by the Deutsche Forschungsgemeinschaft (DFG) through grant DFG Forschergruppe 759 ``The Formation of Planets: The Critical First Growth Phase''.
H.~Beuther acknowledges financial support by the Emmy-Noether-Programm of the DFG through grant BE2578.
The time-consuming three-dimensional simulations have been performed on the pia cluster of the Max Planck Intitute for Astronomy at the computing center of the Max Planck society in Garching.
We acknowledge Andrea Mignone, the main developer of the open source magneto-hydrodynamics code Pluto,
as well as Petros Tzeferacos, who implemented the viscosity tensor into the Pluto code.
We thank our colleague Mario Flock for his dedicated participation in the enhancements of our version of the Pluto MHD code.
We thank Takashi Hosokawa for contributing the stellar evolutionary tracks.
Furthermore, R.~Kuiper thanks Richard Durisen for helpful discussions on disk stability in general.

\bibliographystyle{apj}
\bibliography{Papers}

\begin{thebibliography}{74}
\expandafter\ifx\csname natexlab\endcsname\relax\def\natexlab#1{#1}\fi

\bibitem[{Adams \& Benz(1992)}]{Adams:1992p14918}
Adams, F.~C., \& Benz, W. 1992, in: Complementary Approaches to Double and
  Multiple Star Research, 32, 185

\bibitem[{Adams {et~al.}(1989)Adams, Ruden, \& Shu}]{Adams:1989p14769}
Adams, F.~C., Ruden, S.~P., \& Shu, F.~H. 1989, ApJ, 347, 959

\bibitem[{Anthony \& Carlberg(1988)}]{Anthony:1988p14843}
Anthony, D.~M., \& Carlberg, R.~G. 1988, ApJ, 332, 637

\bibitem[{Balbus(2003)}]{Balbus:2003p11906}
Balbus, S.~A. 2003, ARA{\&}A, 41, 555

\bibitem[{Balbus \& Hawley(1991)}]{Balbus:1991p3799}
Balbus, S.~A., \& Hawley, J.~F. 1991, ApJ, 376, 214

\bibitem[{Bate(1998)}]{Bate:1998p15242}
Bate, M.~R. 1998, ApJ, 508, L95

\bibitem[{Beuther \& Walsh(2008)}]{Beuther:2008p4374}
Beuther, H., \& Walsh, A.~J. 2008, ApJ, 673, L55

\bibitem[{Beuther {et~al.}(2009)Beuther, Walsh, \&
  Longmore}]{Beuther:2009p12913}
Beuther, H., Walsh, A.~J., \& Longmore, S.~N. 2009, ApJS, 184, 366

\bibitem[{Beuther {et~al.}(2005)Beuther, Zhang, Sridharan, \&
  Chen}]{Beuther:2005p1793}
Beuther, H., Zhang, Q., Sridharan, T.~K., \& Chen, Y. 2005, ApJ, 628, 800

\bibitem[{Bodenheimer(1995)}]{Bodenheimer:1995p11927}
Bodenheimer, P. 1995, ARA{\&}A, 33, 199

\bibitem[{Cassen {et~al.}(1981)Cassen, Smith, Miller, \&
  Reynolds}]{Cassen:1981p15165}
Cassen, P., Smith, B.~F., Miller, R.~H., \& Reynolds, R.~T. 1981, Icarus, 48,
  377

\bibitem[{Cesaroni {et~al.}(1997)Cesaroni, Felli, Testi, Walmsley, \&
  Olmi}]{Cesaroni:1997p4541}
Cesaroni, R., Felli, M., Testi, L., Walmsley, C.~M., \& Olmi, L. 1997, A{\&}A,
  325, 725

\bibitem[{Cesaroni {et~al.}(2007)Cesaroni, Galli, Lodato, Walmsley, \&
  Zhang}]{Cesaroni:2007p1870}
Cesaroni, R., Galli, D., Lodato, G., Walmsley, C.~M., \& Zhang, Q. 2007, in:
  Protostars and Planets V, ed. B. Reipurth, D. Jewitt, {\&} K. Keil, 197

\bibitem[{Duschl {et~al.}(2000)Duschl, Strittmatter, \&
  Biermann}]{Duschl:2000p14177}
Duschl, W.~J., Strittmatter, P.~A., \& Biermann, P.~L. 2000, A{\&}A, 357, 1123

\bibitem[{Dzyurkevich {et~al.}(2010)Dzyurkevich, Flock, Turner, Klahr, \&
  Henning}]{Dzyurkevich:2010p15349}
Dzyurkevich, N., Flock, M., Turner, N.~J., Klahr, H., \& Henning, T. 2010,
  A{\&}A, 515, 70

\bibitem[{Enoch {et~al.}(2009)Enoch, Evans, Sargent, \&
  Glenn}]{Enoch:2009p16308}
Enoch, M.~L., Evans, N.~J., Sargent, A.~I., \& Glenn, J. 2009, ApJ, 692, 973

\bibitem[{Flock {et~al.}(2010)Flock, Dzyurkevich, Klahr, \&
  Mignone}]{Flock:2010p15350}
Flock, M., Dzyurkevich, N., Klahr, H., \& Mignone, A. 2010, A{\&}A, 516, 26

\bibitem[{Gammie(2001)}]{Gammie:2001p14271}
Gammie, C.~F. 2001, ApJ, 553, 174

\bibitem[{Goldreich \& Lynden-Bell(1965)}]{Goldreich:1965p14960}
Goldreich, P., \& Lynden-Bell, D. 1965, MNRAS, 130, 125

\bibitem[{Hartmann(1998)}]{Hartmann:1998p14708}
Hartmann, L. 1998, Accretion processes in star formation‎

\bibitem[{Hawley \& Balbus(1991)}]{Hawley:1991p11900}
Hawley, J.~F., \& Balbus, S.~A. 1991, ApJ, 376, 223

\bibitem[{Heemskerk {et~al.}(1992)Heemskerk, Papaloizou, \&
  Savonije}]{Heemskerk:1992p15176}
Heemskerk, M. H.~M., Papaloizou, J. C.~B., \& Savonije, G.~J. 1992, A{\&}A,
  260, 161

\bibitem[{Hosokawa \& Omukai(2009)}]{Hosokawa:2009p12591}
Hosokawa, T., \& Omukai, K. 2009, ApJ, 691, 823

\bibitem[{Isella \& Natta(2005)}]{Isella:2005p3014}
Isella, A., \& Natta, A. 2005, A{\&}A, 438, 899

\bibitem[{Klahr \& Bodenheimer(2003)}]{Klahr:2003p2794}
Klahr, H., \& Bodenheimer, P. 2003, ApJ, 582, 869

\bibitem[{Kley \& Lin(1992)}]{Kley:1992p15153}
Kley, W., \& Lin, D. N.~C. 1992, ApJ, 397, 600

\bibitem[{Kley {et~al.}(1993{\natexlab{a}})Kley, Papaloizou, \&
  Lin}]{Kley:1993p15156}
Kley, W., Papaloizou, J. C.~B., \& Lin, D. N.~C. 1993{\natexlab{a}}, ApJ, 416,
  679

\bibitem[{Kley {et~al.}(1993{\natexlab{b}})Kley, Papaloizou, \&
  Lin}]{Kley:1993p15155}
---. 1993{\natexlab{b}}, ApJ, 409, 739

\bibitem[{Kratter \& Matzner(2006)}]{Kratter:2006p11356}
Kratter, K.~M., \& Matzner, C.~D. 2006, MNRAS, 373, 1563

\bibitem[{Kratter {et~al.}(2008)Kratter, Matzner, \&
  Krumholz}]{Kratter:2008p17135}
Kratter, K.~M., Matzner, C.~D., \& Krumholz, M.~R. 2008, The Astrophysical
  Journal, 681, 375

\bibitem[{Krumholz {et~al.}(2007)Krumholz, Klein, \&
  McKee}]{Krumholz:2007p1380}
Krumholz, M.~R., Klein, R.~I., \& McKee, C.~F. 2007, ApJ, 656, 959

\bibitem[{Krumholz {et~al.}(2009)Krumholz, Klein, McKee, Offner, \&
  Cunningham}]{Krumholz:2009p10975}
Krumholz, M.~R., Klein, R.~I., McKee, C.~F., Offner, S. S.~R., \& Cunningham,
  A.~J. 2009, Science, 323, 754

\bibitem[{Kuiper {et~al.}(2010{\natexlab{a}})Kuiper, Klahr, Beuther, \&
  Henning}]{Kuiper:2010p17191}
Kuiper, R., Klahr, H., Beuther, H., \& Henning, T. 2010{\natexlab{a}}, The
  Astrophysical Journal, 722, 1556

\bibitem[{Kuiper {et~al.}(2010{\natexlab{b}})Kuiper, Klahr, Dullemond, Kley, \&
  Henning}]{Kuiper:2010p12874}
Kuiper, R., Klahr, H., Dullemond, C.~P., Kley, W., \& Henning, T.
  2010{\natexlab{b}}, A{\&}A, 511, 81

\bibitem[{Laor \& Draine(1993)}]{Laor:1993p736}
Laor, A., \& Draine, B.~T. 1993, ApJ, 402, 441

\bibitem[{Laughlin \& Bodenheimer(1994)}]{Laughlin:1994p11930}
Laughlin, G.~P., \& Bodenheimer, P. 1994, ApJ, 436, 335

\bibitem[{Laughlin \& Korchagin(1996)}]{Laughlin:1996p15224}
Laughlin, G.~P., \& Korchagin, V. 1996, ApJ, 460, 855

\bibitem[{Laughlin \& Rozyczka(1996)}]{Laughlin:1996p15239}
Laughlin, G.~P., \& Rozyczka, M. 1996, ApJ, 456, 279

\bibitem[{Levermore \& Pomraning(1981)}]{Levermore:1981p57}
Levermore, C.~D., \& Pomraning, G.~C. 1981, ApJ, 248, 321

\bibitem[{Lin \& Papaloizou(1996)}]{Lin:1996p15022}
Lin, D. N.~C., \& Papaloizou, J. C.~B. 1996, ARA{\&}A, 34, 703

\bibitem[{Lin {et~al.}(1993)Lin, Papaloizou, \& Kley}]{Lin:1993p15154}
Lin, D. N.~C., Papaloizou, J. C.~B., \& Kley, W. 1993, ApJ, 416, 689

\bibitem[{Lin \& Pringle(1990)}]{Lin:1990p14979}
Lin, D. N.~C., \& Pringle, J.~E. 1990, ApJ, 358, 515

\bibitem[{Lodato(2008)}]{Lodato:2008p10897}
Lodato, G. 2008, New Astr. Rev., 52, 21

\bibitem[{Mignone {et~al.}(2007)Mignone, Bodo, Massaglia, Matsakos, Tesileanu,
  Zanni, \& Ferrari}]{Mignone:2007p3421}
Mignone, A., Bodo, G., Massaglia, S., Matsakos, T., Tesileanu, O., Zanni, C.,
  \& Ferrari, A. 2007, ApJS, 170, 228

\bibitem[{Miyama {et~al.}(1994)Miyama, Nakamoto, Kikuchi, Inutsuka, Kobayashi,
  \& Takeuchi}]{Miyama:1994p15169}
Miyama, S.~M., Nakamoto, T., Kikuchi, N., Inutsuka, S.-I., Kobayashi, N., \&
  Takeuchi, T. 1994, in: 1. UNAM-CRAY Supercomputing Workshop: Numerical
  simulations in astrophysics, 305

\bibitem[{Papaloizou \& Lin(1995{\natexlab{a}})}]{Papaloizou:1995p15024}
Papaloizou, J. C.~B., \& Lin, D. N.~C. 1995{\natexlab{a}}, ApJ, 438, 841

\bibitem[{Papaloizou \& Lin(1995{\natexlab{b}})}]{Papaloizou:1995p15025}
---. 1995{\natexlab{b}}, ARA{\&}A, 33, 505

\bibitem[{Papaloizou \& Savonije(1991)}]{Papaloizou:1991p14835}
Papaloizou, J. C.~B., \& Savonije, G.~J. 1991, MNRAS, 248, 353

\bibitem[{Peters {et~al.}(2010{\natexlab{a}})Peters, Banerjee, Klessen, \&
  Low}]{Peters:2010p16326}
Peters, T., Banerjee, R., Klessen, R.~S., \& Low, M.-M.~M. 2010{\natexlab{a}},
  eprint arXiv, 1010, 5905, submitted to ApJ

\bibitem[{Peters {et~al.}(2010{\natexlab{b}})Peters, Banerjee, Klessen, Low,
  Galv{\'a}n-Madrid, \& Keto}]{Peters:2010p12919}
Peters, T., Banerjee, R., Klessen, R.~S., Low, M.-M.~M., Galv{\'a}n-Madrid, R.,
  \& Keto, E.~R. 2010{\natexlab{b}}, ApJ, 711, 1017

\bibitem[{Peters {et~al.}(2010{\natexlab{c}})Peters, Klessen, Low, \&
  Banerjee}]{Peters:2010p17050}
Peters, T., Klessen, R.~S., Low, M.-M.~M., \& Banerjee, R. 2010{\natexlab{c}},
  ApJ, 725, 134

\bibitem[{Rozyczka {et~al.}(1994)Rozyczka, Bodenheimer, \&
  Bell}]{Rozyczka:1994p15157}
Rozyczka, M., Bodenheimer, P., \& Bell, K.~R. 1994, ApJ, 423, 736

\bibitem[{Ruden \& Pollack(1991)}]{Ruden:1991p15061}
Ruden, S.~P., \& Pollack, J.~B. 1991, ApJ, 375, 740

\bibitem[{Shakura \& Sunyaev(1973)}]{Shakura:1973p3060}
Shakura, N.~I., \& Sunyaev, R.~A. 1973, A{\&}A, 24, 337

\bibitem[{Shu {et~al.}(1990)Shu, Tremaine, Adams, \& Ruden}]{Shu:1990p14813}
Shu, F.~H., Tremaine, S., Adams, F.~C., \& Ruden, S.~P. 1990, ApJ, 358, 495

\bibitem[{Strang(1968)}]{Strang:1968p6618}
Strang, G. 1968, SIAM Journal on Numerical Analysis, 5, 506

\bibitem[{Tomley {et~al.}(1991)Tomley, Cassen, \&
  Steiman-Cameron}]{Tomley:1991p14849}
Tomley, L., Cassen, P., \& Steiman-Cameron, T.~Y. 1991, ApJ, 382, 530

\bibitem[{Tomley {et~al.}(1994)Tomley, Steiman-Cameron, \&
  Cassen}]{Tomley:1994p14850}
Tomley, L., Steiman-Cameron, T.~Y., \& Cassen, P. 1994, ApJ, 422, 850

\bibitem[{Toomre(1977)}]{Toomre:1977p14986}
Toomre, A. 1977, ARA{\&}A, 15, 437

\bibitem[{Toomre(1981)}]{Toomre:1981p14967}
---. 1981, in: The structure and evolution of normal galaxies; Proceedings of
  the Advanced Study Institute, 111

\bibitem[{Truelove {et~al.}(1997)Truelove, Klein, McKee, Holliman, Howell, \&
  Greenough}]{Truelove:1997p10742}
Truelove, J.~K., Klein, R.~I., McKee, C.~F., Holliman, J.~H., Howell, L.~H., \&
  Greenough, J.~A. 1997, ApJL, 489, L179

\bibitem[{Tscharnuter \& Boss(1993)}]{Tscharnuter:1993p2109}
Tscharnuter, W.~M., \& Boss, A.~P. 1993, in: Protostars and Planets III, ed. E.
  H. Levy, J. I. Lunine, 921

\bibitem[{Vaidya {et~al.}(2009)Vaidya, Fendt, \& Beuther}]{Vaidya:2009p12873}
Vaidya, B., Fendt, C., \& Beuther, H. 2009, ApJ, 702, 567

\bibitem[{van Boekel {et~al.}(2010)van Boekel, Juhasz, Henning, K{\"o}hler,
  Ratzka, Herbst, Bouwman, \& Kley}]{vanBoekel:2010p14739}
van Boekel, R., Juhasz, A., Henning, T., K{\"o}hler, R., Ratzka, T., Herbst,
  T., Bouwman, J., \& Kley, W. 2010, A{\&}A, 517, 16

\bibitem[{van Leer(1979)}]{vanLeer:1979p5193}
van Leer, B. 1979, JCP, 32, 101

\bibitem[{Vorobyov \& Basu(2006)}]{Vorobyov:2006p16253}
Vorobyov, E.~I., \& Basu, S. 2006, ApJ, 650, 956

\bibitem[{Vorobyov \& Basu(2007)}]{Vorobyov:2007p15725}
---. 2007, MNRAS, 381, 1009

\bibitem[{Vorobyov \& Basu(2009)}]{Vorobyov:2009p16255}
---. 2009, MNRAS, 393, 822

\bibitem[{Vorobyov \& Basu(2010)}]{Vorobyov:2010p16256}
---. 2010, ApJ, 719, 1896

\bibitem[{Yang {et~al.}(1991)Yang, Durisen, Cohl, Imamura, \&
  Toman}]{Yang:1991p11485}
Yang, S., Durisen, R.~H., Cohl, H.~S., Imamura, J.~N., \& Toman, J. 1991,
  Icarus, 91, 14

\bibitem[{Yorke \& Bodenheimer(1999)}]{Yorke:1999p1156}
Yorke, H.~W., \& Bodenheimer, P. 1999, ApJ, 525, 330

\bibitem[{Yorke {et~al.}(1995)Yorke, Bodenheimer, \&
  Laughlin}]{Yorke:1995p1426}
Yorke, H.~W., Bodenheimer, P., \& Laughlin, G.~P. 1995, ApJ, 443, 199

\bibitem[{Yorke \& Sonnhalter(2002)}]{Yorke:2002p1}
Yorke, H.~W., \& Sonnhalter, C. 2002, ApJ, 569, 846

\bibitem[{Zhang {et~al.}(1998)Zhang, Hunter, \& Sridharan}]{Zhang:1998p4666}
Zhang, Q., Hunter, T.~R., \& Sridharan, T.~K. 1998, ApJ, 505, L151

\end{thebibliography}
\end{document}